\documentclass[seceq]{ptptex}

\usepackage{graphics} 

\usepackage{epsfig}
\notypesetlogo        

\makeatletter
\@addtoreset{equation}{section}
\makeatother

\title{%
  \centering
  Complex-Scaling Calculation of Three-Body Resonances
 Using Complex-Range Gaussian Basis Functions \\
 ---  Application to $3\,\!\alpha$ resonaces in $^{12}$\textrm{C} ---
}

\author{
 Shin-Ichi OHTSUBO$^{1}$,
 Yoshihiro FUKUSHIMA$^{1}$,
 Masayasu KAMIMURA$^{2}$  \\
  and Emiko HIYAMA$^{2}$
}

\inst{
$^{1}$Department of Applied Physics, Fukuoka University, 
Fukuoka 814-0180, Japan\\
$^{2}$RIKEN Nishina Center, RIKEN, Wako 351-0198, Japan
}

\abst{
We propose to use the complex-range Gaussian basis functions,
$\{r^l\:e^{- ( 1 \pm i\, \omega )(r/r_n)^2 }
Y_{lm}(\widehat{\mathbf{r}}); \\
r_n$ in a geometric progression\}, 
in the calculation of three-body resonances
with the complex-scaling method (CSM) in which  
use is often made of the real-range Gaussian basis functions,
$\{r^l\:e^{-(r/r_n)^2 }Y_{lm}(\widehat{\mathbf{r}})\}$,
that are suitable for describing the short-distance structure and
the asymptotic decaying behavior of few-body systems.
The former basis set is  more powerful   
than the latter when describing the resonant and nonresonant
continuum states
with highly oscillating amplitude at large scaling angles $\theta$. 
We applied the new basis
functions to the CSM calculation of the $3 \alpha$ resonances with 
$J=0^+, 2^+$ and $4^+$ in $^{12}$C. 
The eigenvalue distribution
of the complex scaled Hamiltonian becomes more precise 
and the maximum scaling angle becomes drastically larger 
($\theta_{\textrm{max}}=16^\circ \to 36^\circ$)
than those given by the use of the real-range Gaussians.
Owing to these advantages, we were able to confirm the prediction by
Kurokawa and Kat\={o} [Phys. Rev. C \textbf{71}, 021301 (2005)] 
on the appearance of the new broad $0^+_3$ state; we show it as
an explicit resonance pole isolated from the $3\alpha$ continuum.
}

\begin{document}

\maketitle

\section{Introduction}
\label{sect:Introduction}

The complex scaling method
 (CSM)\cite{CSM-ref1,CSM-ref2,CSM-ref3,Ho,Moiseyev}
is a very powerful tool to investigate
resonances in quantum many-body systems.
Application of the CSM to the nuclear physics problems
are extensively reviewed in Ref.~\citen{CSM-review} and references therein.
In the CSM the resonance parameters can be obtained by
using only $L^2$ (bound state type) wave functions and 
without the explicit scattering
calculations or without the use of the continuum wave functions;
namely, 
the energy $E_r$ and 
the decay width $\varGamma$ of a resonance
can be obtained 
by solving the eigenvalue problem for the complex scaled 
Schr\"{o}dinger equation 
with a scaling angle~$\theta$, $[H(\theta) -E(\theta)] \varPsi(\theta)=0$,
where $\varPsi(\theta)$ are expanded in terms of only $L^2$ integrable
many-body basis functions.
   
In the CSM, there is a limitation of the scaling angle $\theta$ 
due to the analyticity of the Hamiltonian. 
Furthermore, in practical calculations, 
one often meets a difficulty in solving 
resonant states with large decay width since the complex scaled Hamiltonian
is diagonalized with a limited number of basis functions.
A set of real-range Gaussians,
$\{r^l\:e^{-(r/r_n)^2} Y_{lm}(\widehat{\mathbf{r}}); \: 
r_n$ in a geometric progression\}~\cite{Kamimura88,Kameyama89,
Hiyama03,Hiyama2012ptep},
are often employed as the basis functions.
But, it is difficult for the basis set to describe highly 
oscillating wave functions that appears in the CSM 
when the scaling angle becomes large; the overlap matrix 
of the basis set becomes easily ill-conditioned
when the number of basis functions is rather large. 

Thus, one of the purposes of the present work is
to propose the use of the complex-range Gaussian basis functions,
$\{r^l\:e^{- ( 1 \pm i\, \omega )(r/r_n)^2 } 
Y_{lm}(\widehat{\mathbf{r}})\}$~\cite{Hiyama03,Hiyama2012ptep}, 
in the CSM calculation of three-body resonances
so as to overcome the above difficulty for the large $\theta$.
Owing to the oscillating component,
the space of the new function set becomes much larger and
the overlap matrix hardly becomes ill-conditioned.
This improves the quality of the CSM calculation significantly
and increases the possible scaling angle drastically. 

One of the most intensively studied nuclei using the three-body CSM is
$^{12}$C nucleus as the $3\alpha$-cluster system.
The CSM has especially been useful to investigate the
$3\alpha$  resonance structure in its  excited states.
Such CSM studies of $^{12}$C
are reviewed in Refs.~\citen{Kurokawa1} and \citen{Kurokawa2} 
and references therein.
Among the studies, Kurokawa and Kat\={o}~\cite{Kurokawa1,Kurokawa2}
succeeded in thoroughly calculating 
the energies and decay widths of the $3\alpha$ 
resonant states in $^{12}$C with the total angular momentum $J=0$ to 5.

\begin{table}[!ht]
\centering
\caption{ 
New results of  the recent calculation~\cite{Kurokawa1,Kurokawa2}
and experiment~\cite{Itoh2011} 
on the $0^+_3$, $0^+_4$ and $2^+_2$ states in $^{12}$C.
The excitation energies $(E_x)$ and  decay widths $(\varGamma)$ are
given in \mbox{MeV}. See also the footnote below.
}
\vspace{5pt}
\label{tab:Kurokwa-Exp}
\begin{tabular}{c c r@{.}l r@{.}l c r@{.}l@{$\,\pm\,$}r@{.}r
                                    r@{.}l@{$\,\pm\,$}r@{.}r }
\hline  \\ [-6pt]
  $^{12}$C    & \phantom{??}
              & \multicolumn{4}{c}{CAL~\cite{Kurokawa1,Kurokawa2}}
              & \phantom{??}
              & \multicolumn{8}{c}{EXP~\cite{Itoh2011}}   \\ [3pt]
\cline{3-6} \cline{8-15}  \\ [-8pt]
  $J^{\pi}$   &{}
              & \multicolumn{2}{c}{$E_{x}$}
              & \multicolumn{2}{c}{$\varGamma$} &{}
              & \multicolumn{4}{c}{$E_{x}$}
              & \multicolumn{4}{c}{$\varGamma$}    \\ [2pt]
\hline  \\ [-6pt]
  $0^{+}_{3}$ &{}&   8&95   &   1&48   &{}&  9&04  &  0&09
                                          &  1&45  &  0&18  \\
  $0^{+}_{4}$ &{}&  11&87   &   1&1    &{}& 10&56  &  0&06
                                          &  1&42  &  0&08  \\
  $2^{+}_{2}$ &{}&   9&57   &   1&1    &{}&  9&84  &  0&06
                                          &  1&01  &  0&15  \\ [3pt]
\hline
\end{tabular}
\end{table}

In Table~\ref{tab:Kurokwa-Exp}, it is interesting to see the recently accomplished
reasonable agreement between the results by 
the calculation~\cite{Kurokawa1,Kurokawa2}
and the observation~\cite{Itoh2011} 
on the $0^+_3$, $0^+_4$ and $2^+_2$ states. 
Especially, Kurokawa-Kat\={o}'s prediction of the new $0^+_3$ state having
a large width is of importance to understand the
new experimental results 
in Ref.~\citen{Itoh2011}.\footnote{ 
As for the $0^+$ data in Fig.~8(a) in Ref.~\citen{Itoh2011}, 
we employ the interpretation by the authors  that 
there are two  $0^+$ resonance peaks as summerized in Table~\ref{tab:Kurokwa-Exp}
and that the resonances may correspond respectively 
to the $0^+_3$ and $0^+_4$ states described
in Ref.~\citen{Kurokawa1}. The authors showed another
interpretation to regard the peaks as a single peak  at
$E_x=9.93 \pm 0.03$ \mbox{MeV} with a width of $2.71 \pm 0.08$ \mbox{MeV}.
}
Their CSM calculation was performed at the scaling angle of $16^\circ$
that is the largest angle available in the 
calculation using the real-range Gaussian basis.
Since the angle is not enough for separating 
the low-lying broad $0^+_3$ resonance from the continuum eigenvalues, 
they made an extrapolation
by applying the method of analytic continuation of the coupling 
constant (ACCC)~\cite{ACCC,ACCC-book}
combined with the CSM (ACCC+CSM)~\cite{ACCC-Aoyama}
in order to derive the complex energy of the $0^+_3$ resonance.
But, Arai~\cite{Arai} reported that, the $0^+_3$ resonance obtained in
Ref.~\citen{Kurokawa1} is missing from
his calculation based on
the microscopic $R$-matrix method for the $^8{\textrm{Be}}(0^+, 2^+, 4^+)
+ \alpha$ two-body scattering problem.  However, in the 
calculation in Ref.~\citen{Arai},
the $^8$Be was described with the bound-state approximation
employing only four different tempered Gaussian 
functions for the $\alpha$-$\alpha$ relative motion.

Thus, the second purpose of the present paper is to apply the 
CSM with the complex-range Gaussian basis functions to the
$3\alpha$ resonances in $^{12}$C and examine the
results in Refs.~\citen{Kurokawa1} and \citen{Kurokawa2}.
Since the function space of the present basis set is very large,
we have the following advantages:
i) the distribution of the eigenvalues of the complex scaled Hamiltonian
becomes much more precise than those obtained in the literature,
and ii) the scaling angle is drastically increased from $\theta=16^\circ$ 
up to $36^\circ$ that is enough large to separate explicitly 
the $0^+_3$ resonance pole from the $3\alpha$ continuum eigenvalues.

The present paper is organized as follows: in Section \ref{sect:Method},
we introduce the complex-range Gaussian basis functions 
and incorporate them into the framework of CSM.
In Section \ref{sect:Application}, we apply the method to the $3\alpha$ resonances
in $^{12}$C and compare the result with that obtained 
in Refs.~\citen{Kurokawa1} and \citen{Kurokawa2}.  
Summary is given in Section \ref{sect:Summary}.

\section{Method}
\label{sect:Method}

\subsection{Three-body complex scaling method}

In many cases of the CSM studies, the three-body wave function 
is expanded in terms of the real-range 
Gaussian basis set with ranges in
a geometric progression. In this work, we propose to use 
the complex-range Gaussians in the three-body CSM calculations.

\begin{figure}[!ht]
\centering
  \includegraphics[height=2.8cm,keepaspectratio,clip]{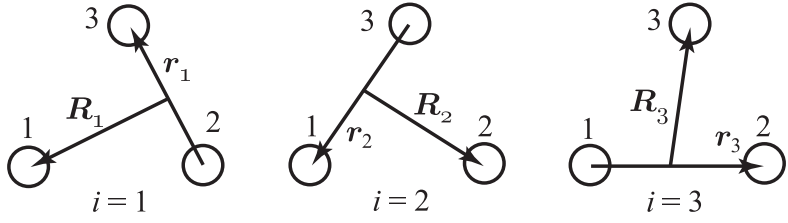}
\caption{
Three sets of the Jacobi coordinates for three $\alpha$ particles.
}%
\label{fig:fig1-jacobi}
\end{figure}

We explain it, as an example,  taking the case of 
$^{12}{\textrm{C}}(=\alpha+\alpha+\alpha)$ on the basis of 
the orthogonality condition model (OCM)~\cite{Saito69} 
for the $3 \alpha$ system.
The extension from the real-range Gaussian to the complex-range ones
in other three-body systems is straightforward.
We take all the three sets of 
Jacobi coordinates (Fig.~\ref{fig:fig1-jacobi}),
$\textbf{r}_{1}=\textbf{x}_2 -\textbf{x}_3$ and 
$\textbf{R}_{1}=\textbf{x}_1 -\frac{1}{2}(\textbf{x}_2 + \textbf{x}_3)$ 
and cyclically for $(\textbf{r}_2, \textbf{R}_2)$ and 
$(\textbf{r}_3, \textbf{ R}_3)$,
$\textbf{x}_i$ being the position vector of $i$th particle. 

The Hamiltonian is written as
\begin{equation}
\label{eq:Hamiltonian}
  H= \sum_{i=1}^3 t_i  - T_{\textrm{G}} + 
  \sum_{i=1}^3 V_{\alpha \alpha}(r_{i})
  + V_{3\alpha}(r_1, r_2, r_3) + V_{\textrm{Pauli}}.
\end{equation}
The operators $t_i$ and $T_{\textrm{G}}$ stand for the kinetic energies of 
$i$th $\alpha$ particle and the center of mass motion, respectively. 
$V_{\alpha \alpha}$ is the $\alpha$-$\alpha$ potential
and $V_{\textrm{Pauli}}$ is a pseudo potential representing 
the Pauli principle between $\alpha$ clusters.
The $3 \alpha$ potential $V_{3\alpha}$ is introduced if necessary.
These potentials are explained in Subsection \ref{subsect:Interaction}.

In the CSM, the radial coordinates are transformed by
\begin{equation}
r_i \to r_i \,e^{i \theta}, \qquad  
R_i \to R_i \,e^{i \theta}. 
\end{equation}
The transformed Hamiltonian is denoted by $H(\theta)$.
We solve the equation
\begin{equation}
\label{eq:variation}
  \left[\, H(\theta)-E(\theta) \,\right] \varPsi(\theta)=0
\end{equation}
by expanding $\varPsi(\theta)$ in terms of the totally symmetric
$L^2$-integrable three-body basis functions 
$\{ \varPsi_\gamma ; \gamma=1,\ldots , \gamma_{\textrm{max}} \}$ :
\begin{equation}
\label{eq:base-expand}
 \varPsi(\theta)=
 \sum_{\gamma=1}^{\gamma_{\textrm{max}}} C_\gamma(\theta) \varPsi_\gamma.
\end{equation}
The complex eigenenergies and the expansion coefficients are 
determined by
\begin{equation}
\label{eq:eigen-equ}
  \sum_{\gamma'=1}^{\gamma_{\textrm{max}}}
    \left[\,H_{\gamma \gamma'}(\theta)-E(\theta)N_{\gamma \gamma'} \,\right]
C_{\gamma'}(\theta) =0 \qquad (\gamma=1,\ldots,\gamma_{\textrm{max}}),
\end{equation}
where the overlap and Hamiltonian matrix elements are respectively
written as
\begin{equation}
N_{\gamma \gamma'}=\langle \varPsi_\gamma | \varPsi_{\gamma'} \rangle, 
\end{equation}
and
\begin{equation}
H_{\gamma \gamma'}(\theta)=\langle \varPsi_\gamma | H(\theta) 
| \varPsi_{\gamma'} \rangle.
\end{equation}
The complex resonance energy is given,  independently 
of $\theta$ in principle, by
\begin{equation}
E_{res} =E_r - \frac{i\varGamma}{2},
\end{equation}
where $E_r$ is the resonance energy with respect to the
$3\alpha$ breakup threshold and $\varGamma$ is the total decay width.

The symmetric three-body basis 
functions $\varPsi_\gamma$ in (\ref{eq:base-expand}) is  written as
\begin{eqnarray}
 \varPsi_\gamma =
    \varPhi_\gamma( \textbf{r}_1, \textbf{R}_1) 
 +  \varPhi_\gamma( \textbf{r}_2, \textbf{R}_2) 
 +  \varPhi_\gamma( \textbf{r}_3, \textbf{R}_3). \qquad 
\end{eqnarray}
We express each 
$\varPhi_\gamma ( \textbf{r}_i, \textbf{R}_i)$  
as a product of a function of $\textbf{r}_i$ and 
that of $\textbf{R}_i$:
\begin{equation}
\label{eq:Product-Func}
 \varPhi_\gamma( \textbf{r}_i, \textbf{R}_i)=
 \phi_{n l}(r_i)
        \psi_{N L}(R_i) 
\left[ Y_{l}({\widehat{ \mathbf{r}}_i}) 
Y_{L}({\widehat{\mathbf{R} }_i})
  \right]_{J M},
\end{equation}
where $\gamma$ specifies a set of
quantum numbers
\begin{equation}
  \gamma \equiv \{n l, N L ,JM \}. 
\end{equation}
$J$ is the total angular momentum  and
$M$ is its $z$-component.

\subsection{Real-range Gaussian basis functions}

According to the Gaussian expansion method (GEM)~\cite{Kamimura88,
Kameyama89,Hiyama03,Hiyama2012ptep}, we take the radial shape of
$\phi_{n l}(r)$ and $\psi_{N L}(r)$ in (\ref{eq:Product-Func})
as follows:
\begin{eqnarray}
\label{eq:RG-Base1}
  \phi_{n l}(r) &=&r^{l}\:e^{-(r/r_{n})^2},\: \;   \\
\label{eq:RG-Base2}
  \psi_{N L}(R)&=&R^{L}\:e^{-(R/R_{N})^2}, \: \;   
\end{eqnarray}
where   normalization constants  are
omitted for simplicity. 
The GEM recommends to set the Gaussian ranges
in a geometric progression:   
\begin{eqnarray}
\label{eq:Geome-Progr-1}
  r_{n}&=& r_1\, a^{n-1}
    \quad \quad (n=1, \ldots , n_{\textrm{max}})\;,  \\
\label{eq:Geome-Progr-2}
  R_{N}&=&R_1\, A^{N-1}
    \quad \:(N=1, \ldots , N_{\textrm{max}})\;.  
\end{eqnarray}
This greatly reduces 
the nonlinear parameters to be optimized.
We designate a set of
the geometric sequence  
by $\{ n_{\textrm{max}},\, r_1, \,
r_{n_{\textrm{max}}} \}$  instead of   
$\{ n_{\textrm{max}},\, r_1, \,a  \}$ 
and similarly for  $\{ N_{\textrm{max}},\, R_1, \,
R_{N_{\textrm{max}}} \}$ , which is more convenient
for consideration of  the spatial distribution of the basis set.

The basis set \{$\phi_{n l}; n=1,\ldots, n_{\rm max}$\}
has the following properties:
i) They range from very compact to very diffuse, more densely 
in the inner region than in the outer one.
While the basis functions with small ranges are responsible for describing 
the short-range structure of the system, the basis 
with longest-range parameters is for the asymptotic behavior. 
ii) After multiplication by normalization
constants for $\langle \phi_{n l} \,|\, \phi_{n l} \rangle =1$,
they have the relation
\begin{equation}
 \langle \phi_{n \, l} \,|\, \phi_{n+k \, l} \rangle =
\left( \frac{2a^k}{1+a^{2k}} \right)^{l+3/2} ,
\end{equation}
which shows that the overlap with the $k$th neighbor
is \textit{independent} of $n$, decreasing gradually with increasing $k$.
We thus expect that the coupling among all the  basis functions
takes place smoothly and coherently so as to
describe properly both the short-range structure and
long-range decaying behavior simultaneously.

We note that a single Gaussian $e^{-(r/r_n)^2}$ 
decays quickly as $r$ increases, 
but appropriate
superposition of many Gaussians can decay even  exponentially 
with increasing $r$ up to a
sufficiently large $r$. Good examples are shown 
in Figs. 3 and 4 in Ref.~\citen{Hiyama03}.  
 
\subsection{Complex-range Gaussian basis functions}

For the precise CSM calculations of 
three-body systems, however,
we improve the Gaussian shape 
to have more sophisticated (but still tractable) radial dependence. 
This is because that the wave function in CSM
becomes more oscillatory as the scaling angle $\theta$ increases.
But, the superposition of the real-range Gaussians is
difficult to accurately describe oscillatory functions having 
several nodes. 

In the GEM in Ref.~\citen{Hiyama03}, it was proposed to improve 
the Gaussian shape   
by introducing the \textit{complex}  range instead of the real one: 
\begin{eqnarray}
\label{eq:CG-Base1-p}
  \phi_{nl}^{(+\omega)}(r) &=&
   r^l\:e^{- ( 1 + i\, \omega )(r/r_n)^2 } ,\\
\label{eq:CG-Base1-m}
  \phi_{nl}^{(-\omega)}(r) &=&
   r^l\:e^{- ( 1 - i\, \omega )(r/r_n)^2 }, 
\end{eqnarray}
and 
\begin{eqnarray}
\label{eq:CG-Base2-p}
  \psi_{NL}^{(+\omega)}(R) &=&
    R^L\:e^{- ( 1 + i\, \omega )(R/R_N)^2 } ,\\
\label{eq:CG-Base2-m}
  \psi_{NL}^{(-\omega)}(R) &=&
    R^L\:e^{- ( 1 - i\, \omega )(R/R_N)^2 } ,
\end{eqnarray}
where the ranges $r_n$ and $R_N$ are given by
(\ref{eq:Geome-Progr-1}) and (\ref{eq:Geome-Progr-2}), respectively.
Using  the above complex conjugate pairs,
$\phi_{nl}^{( \pm \omega)}(r)$ and $\psi_{NL}^{( \pm \omega)}(R)$,
we can construct
equivalent  sets of \textit{real} basis functions:
\begin{eqnarray}
\label{eq:CG-Base1-cos}
  \phi_{nl}^{({\cos})}(r) &=&  
   r^l\:e^{-(r/r_n)^2}\, {\cos}\, \omega (r/r_n)^2, \\
\label{eq:CG-Base1-sin}
  \phi_{nl}^{({\sin})}(r) &=& 
    r^l\:e^{-(r/r_n)^2}\, {\sin}\, \omega (r/r_n)^2, 
\end{eqnarray}
and 
\begin{eqnarray}
\label{eq:CG-Base2-cos}
  \psi_{NL}^{({\cos})}(R) &=&  
    R^L\:e^{-(R/R_N)^2}\, {\cos}\, \omega (R/R_N)^2, \\
\label{eq:CG-Base2-sin}
  \psi_{NL}^{({\sin})}(R) &=& 
    R^L\:e^{-(R/R_N)^2}\, {\sin}\, \omega (R/R_N)^2 .
\end{eqnarray}
In the present CSM calculation of $^{12}$C, 
the former set, (\ref{eq:CG-Base1-p})-(\ref{eq:CG-Base2-m}),
is employed\footnote{  
We made the same calculation employing the latter set,
 (\ref{eq:CG-Base1-cos})-(\ref{eq:CG-Base2-sin})
to crosscheck the computation programs and obtained, 
as a matter of course, the same result. 
}  
and the three-body basis function 
$\varPhi_\gamma( \textbf{r}_i, \textbf{R}_i)$ of (\ref{eq:Product-Func})
is replaced by 
\begin{equation}
 \varPhi_\gamma(\textbf{r}_i, \textbf{R}_i)=
\phi^{(\pm \omega)}_{n l}(r)
     \,   \psi^{(\pm \omega)}_{N L}(R_i) 
\left[ Y_{l}({\widehat{\mathbf{r}}_i}) 
Y_{L}({\widehat{\mathbf{R}}_i})
  \right]_{J M} ,
\end{equation}
with $\gamma$  specifying  a set 
\begin{equation}
  \gamma \equiv \{ \pm \omega, n l;\: \pm \omega, N L; \:JM \},
\end{equation}
where one can take different $\omega$'s between 
$\phi(r)$ and $\psi(R)$ although it was not necessary in the present 
$3\alpha$ CSM calculation.

The new basis functions, (\ref{eq:CG-Base1-p})-(\ref{eq:CG-Base2-sin}),
apparently extend the function space from 
the old ones, (\ref{eq:RG-Base1}) and (\ref{eq:RG-Base2}),
since they have the oscillating components; 
their applications are seen
in Refs.~\citen{Hiyama03, Matsumoto03,
Nakada,Kamimura09, Hiyama2012, Hiyama2012ptep}. 
Note that the computation programming 
is almost the same as that for (\ref{eq:RG-Base1}) and (\ref{eq:RG-Base2})
although some of real variables are changed to complex ones.

\begin{figure}[!ht]
\begin{minipage}{0.45\textwidth}
\centering
\vspace*{-3mm}
\hspace*{-3mm}
  \includegraphics[height=5.6cm,width=6.7cm,clip]{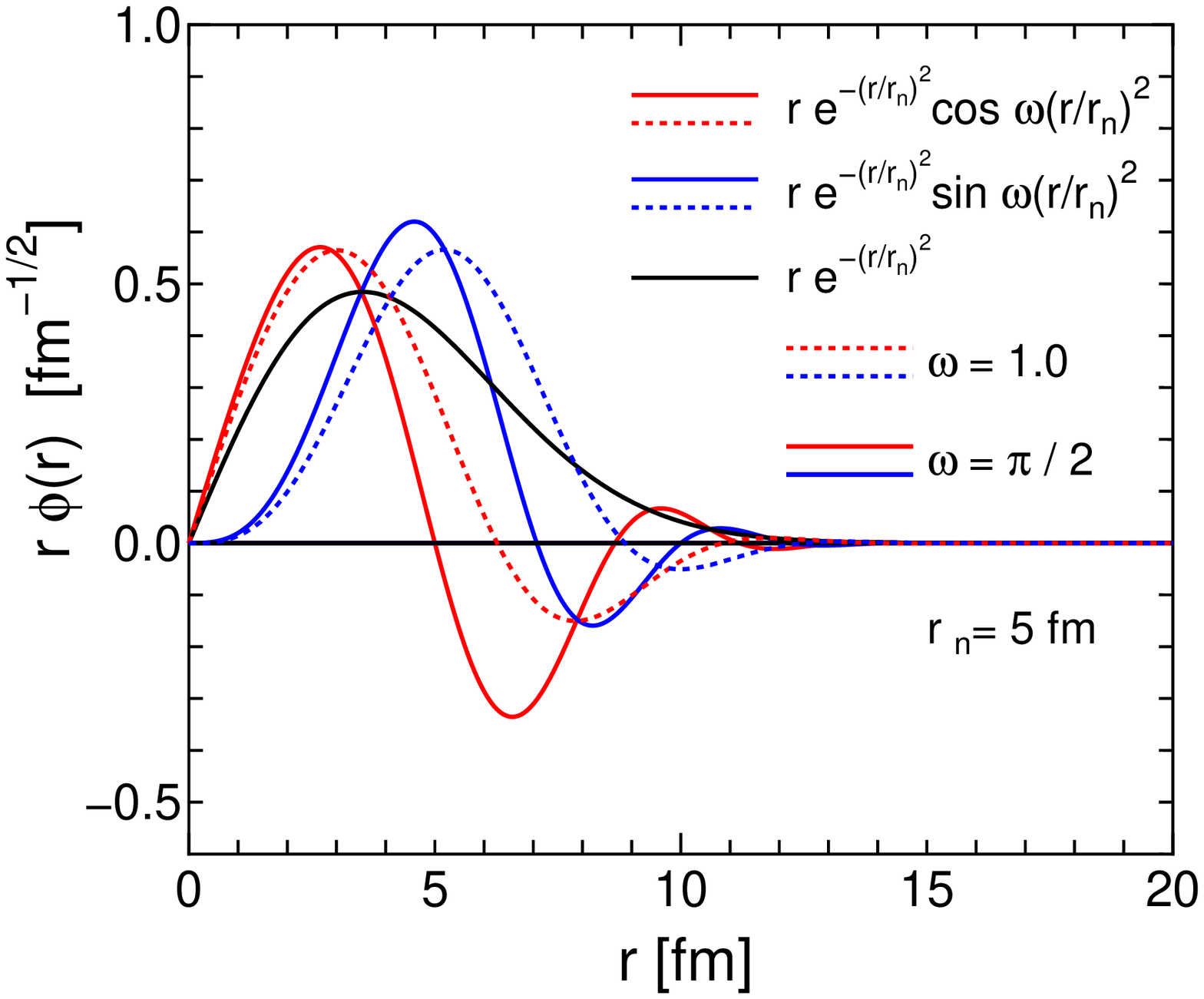}%
\caption{
An example of the $l=0$ complex-range Gaussian basis functions  
presented in the form of Eqs.~(\ref{eq:CG-Base1-cos}) and (\ref{eq:CG-Base1-sin})
with \mbox{$r_n=5$ fm} and $\omega=1.0$ and $\pi/2$.
The functions are normalized to unity. 
}%
\label{fig:fig2-cgauss}%
\end{minipage}%
%
\hspace*{5mm}
\begin{minipage}{0.5\textwidth}
\centering
  \includegraphics[height=5.6cm,width=6.7cm,clip]{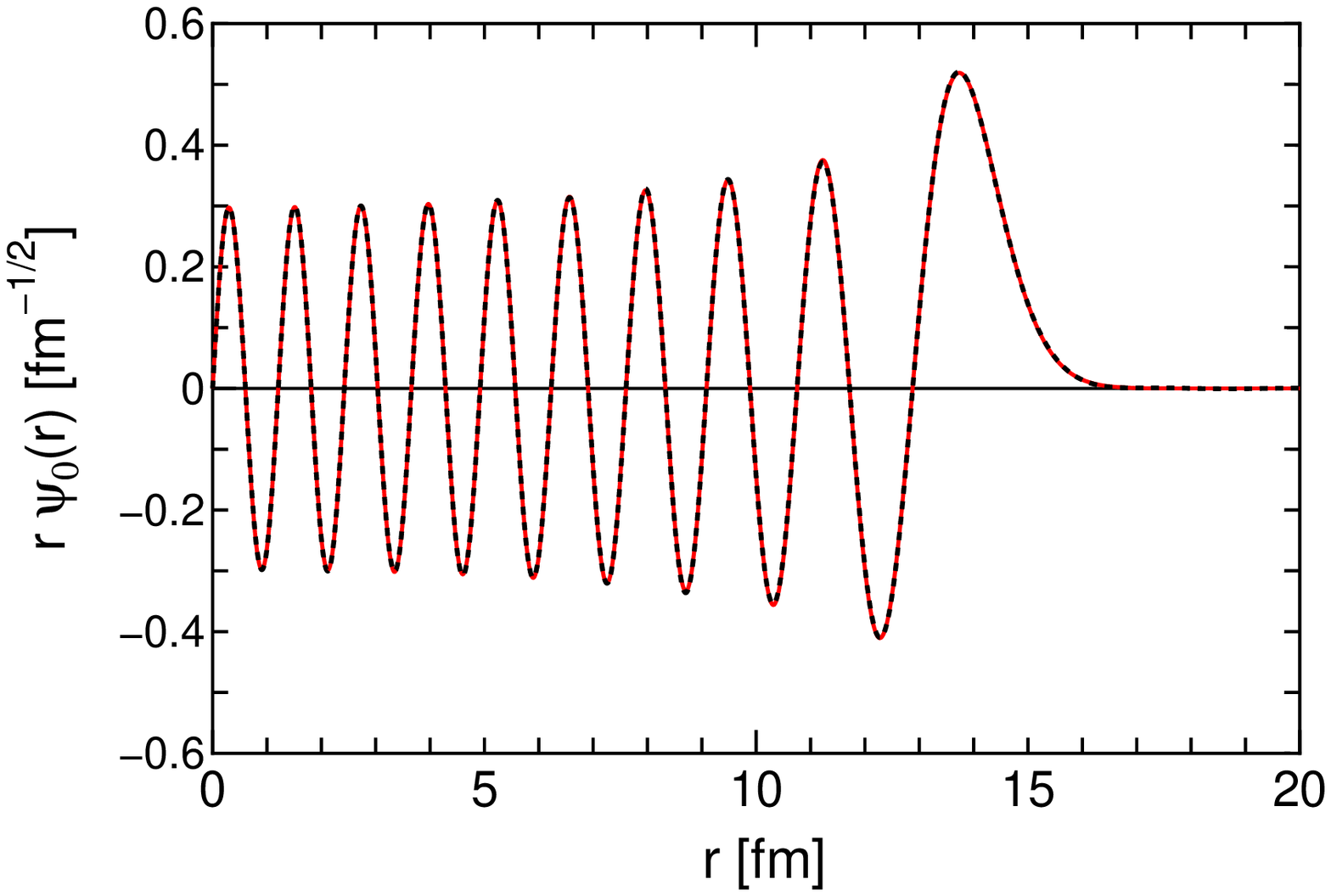}%
\caption{
  Wave function of the $l=0$ \mbox{36-quanta} state  
  for the HO potential using 32 complex-range Gaussians
 (black dotted line) . Deviation from 
 the exact one  (red solid line) is everywhere less than 0.001  
in the unit of the vertical axis.
}%
\label{fig:fig3-howave}%
\end{minipage}
\end{figure}

In order to compare visually the real-range and complex-range Gaussians,
we plot, in Fig. \ref{fig:fig2-cgauss},  
$\phi_{nl}(r)$ of (\ref{eq:RG-Base1}),
$\phi_{nl}^{(\cos)}(r)$ of (\ref{eq:CG-Base1-cos})
and $\phi_{nl}^{(\sin)}(r)$ of (\ref{eq:CG-Base1-sin}) with $l=0$, 
$r_n=5 $ \mbox{fm} and $\omega=1.0$ and $\pi/2$.
A good test of the use of complex-range Gaussians
is to calculate the wave functions of 
highly excited states in a three-dimensional harmonic
oscillator (HO) potential.  We calculate the $l=0$ neutron wave function
in the potential with $\hbar \omega=15.0 $ \mbox{MeV}.
The wave function, $\varPsi_l$, is expanded in terms of
totally 32 basis functions of 
(\ref{eq:CG-Base1-cos}) and (\ref{eq:CG-Base1-sin}) as
\begin{equation}
\varPsi_l(r) = \sum_{n=1}^{n_{\textrm{max}}}
\left[ c_{n l}^{(\cos)} \phi^{(\cos)}_{n l} (r) 
     + c_{n l}^{(\sin)} \phi^{(\sin)}_{n l} (r) \right]
\end{equation}
with $n_{\textrm{max}}=\frac{32}{2}$ and $\omega=1$,
and in terms of 32 real-range Gaussians (\ref{eq:RG-Base1}) as
\begin{equation}
\varPsi_l(r) = \sum_{n=1}^{n_{\textrm{max}}} 
 c_{n l}\, \phi_{n l} (r) 
\end{equation}
with $n_{\textrm{max}}=32$.
The expansion coefficients and the eigenenergies are obtained
by diagonalizing the Hamiltonian in the space.
Optimized nonlinear parameters of the complex-range
Gaussian set are 
\{$n_{\textrm{max}}\!\!=\!\!\frac{32}{2},r_1\!\!=\!\!1.4 \, \textrm{fm}, 
\;r_{n_{\textrm{max}}}=7.1 \,\textrm{fm},
\; \omega=1.0$\}
and those for the real-range Gaussians 
are $\{ n_{\textrm{max}}\!\!=32, r_1=0.6 \, \textrm{fm}, 
\;r_{n_{\textrm{max}}}=16.0 \,\textrm{fm} \} $.
The  range parameters are given by
round numbers, but further 
optimization does not give any significant change to the result.

In Table~\ref{tab:Complex-Gauss-Eig}, 
the calculated energies (in the number of quanta) are compared 
with the exact ones.
The result with the complex-range Gaussians is much better than that 
with the real-range Gaussians  especially in the highly  
oscillatory  states as is expected.
It is to be noted that the both 
cases have the same number of basis functions and 
that the overlap matrix of the real-range Gaussian
basis set becomes heavily ill-conditioned when 
the range parameters \{$r_1, r_{n_{\textrm{max}}}\}$  
are taken to be the same as those of the complex-range Gaussian set.
Extension of the function space due to making the range parameters complex
is much more effective than the simple extension having two
times more functions in the real-range Gaussian set.

\begin{table}[!ht]
\centering
\caption{
 Energies (number of quanta) of the highly excited $l\!=\!0$ states of 
the HO potential calculated using  32 complex-range Gaussians
  with 
\{$n_{\textrm{max}}\!\!=\!\!\frac{32}{2},r_1\!\!=\!\!1.4 \, \textrm{fm}, 
\;r_{n_{\textrm{max}}}=7.1 \,\textrm{fm},
\; \omega=1.0$\} and  32 real-range
Gaussians with 
\{$n_{\textrm{max}}\!\!=\!\!32, r_1\!\!=\!\!0.6 \, \textrm{fm}, 
\;r_{n_{\textrm{max}}}=16.0 \,\textrm{fm}$\}.
}
\begin{tabular}{lllll}
\noalign{\vskip 0.2 true cm} 
\hline
\vspace{-3 mm} \\
 \multicolumn{2}{c}{$\;\;$Exact$\;\;$}& Complex-range & &real-range  \\
& & Gaussians && Gaussians  \\
\hline 
\vspace{-3 mm} \\
&$\;\: 0$ & $\;\;10^{-13}$ & & $10^{-14}$  \\
&12 &  12.000000003     & &    12.00003 \\
&16 &  16.00000005      & &   16.003  \\
&20 &  20.0000005       & &   20.1  \\
&24 &  24.000004        & &   24.6   \\
&28 &  28.00002         & &   30.1   \\
&32 &  32.0003          & &   36.8   \\
&36 &  36.0004          & &   45.2  \\
&40 &  40.04            & &   56.7\\
&44 &  44.2             & &   71.7  \\
\noalign{\vskip 0.1 true cm} 
\hline
\noalign{\vskip 0.3 true cm} 
\end{tabular}
\label{tab:Complex-Gauss-Eig}
\end{table}

In Fig.~\ref{fig:fig3-howave},
wave function of the \mbox{36-quanta} state obtained with the 
complex-range Gaussians is compared with the exact one.  
The two curves for those wave functions overlap to each other everywhere;
the difference is less than 0.001 in the unit of the vertical axis.

We thus expect that  
use of the new basis set (2.17)-(2.20) in three-body CSM calculations
well describes the highly 
oscillating wave functions of  both
the resonant and  nonresonant continuum states 
even when the scaling angle becomes rather large.

\section{Application to $3\alpha$ resonances in $^{12}$C}
\label{sect:Application}

\subsection{Interaction of the $3 \alpha$ system}
\label{subsect:Interaction}

We take the same model and interaction as those in 
Refs.~\citen{Kurokawa1} and \citen{Kurokawa2}.
The potential $V_{\alpha \alpha}$ is constructed by folding 
the effective $N$-$N$ interaction
by Schmid-Wildermuth~\cite{SW-force} and the Coulomb potential
into the density of the $\alpha$ cluster having the $(0s)^4$ configuration.
In Refs.~\citen{Kurokawa1} and \citen{Kurokawa2},
the $V_{\alpha \alpha}$ is 
adjusted to reproduce the experimental phase shift of the
$\alpha$-$\alpha$ system by taking 1.03$\times V_{\alpha \alpha}$.

The Pauli principle between $\alpha$ clusters
is taken into account by the OCM~\cite{Saito69}.
The OCM projection operator \cite{Kukulin84}, $V_{\textrm{Pauli}}$,
in the Hamiltonian~(\ref{eq:Hamiltonian}) is written by
\begin{equation}
\label{eq:V-Pauli}
  V_{\textrm{Pauli}}=\lim_{\lambda\to\infty}  \lambda 
   \sum_f | f \rangle \langle f| \:,
\end {equation}
which rules out the 
Pauli-forbidden $\alpha$-$\alpha$ relative
states $(f=0S, 1S, 0D)$ 
from the three-body  wave function.
In this work, we take $\lambda=10^5$ \mbox{MeV}.

Since use of the $2 \alpha$ potential $V_{\alpha \alpha}$ together with the
Pauli potential $V_{\textrm{Pauli}}$ makes the energies of
the ground-rotational-band states $(0^+_1, 2^+_1, 4^+_1)$ 
lower than the observed values,
the repulsive $3 \alpha$ potential $V_{3 \alpha}$ in (\ref{eq:Hamiltonian})
is introduced in Refs.~\citen{Kurokawa1} and \citen{Kurokawa2}
phenomenologically in the form
\begin{equation}
\label{eq:V-alpha-alpha}
  V_{3\alpha}(r_1,r_2,r_3)= V_{3\alpha}^{J^\pi} 
   \exp \left[\, -\mu(r_1^2+r_2^2+r_3^2) \,\right],
     \qquad \mu=0.15 \:\textrm{fm}^{-2},
\end{equation}
where 
$V_{3\alpha}^{0^+}=31.7$ \mbox{MeV},
$V_{3\alpha}^{2^+}=63.0$ \mbox{MeV} and 
$V_{3\alpha}^{4^+}=150.0$ \mbox{MeV} are employed dependently on the total
angular momentum $J=0^+, 2^+$ and $4^+$, respectively.

\subsection{The $0^+$ resonances}

\begin{figure}[!ht]
\centering
  \includegraphics[width=6.5cm,height=6.5cm,clip]{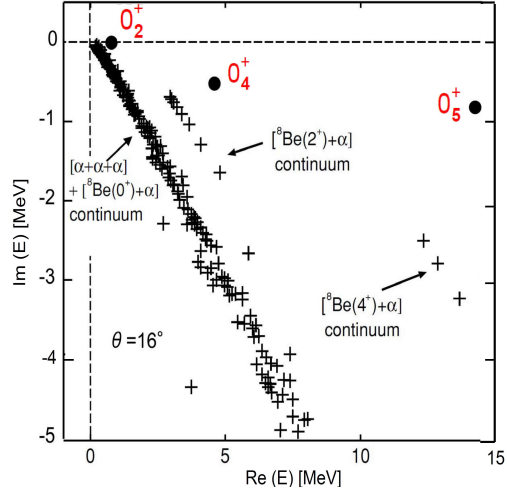}%
\caption{
The $0^+$ eigenvalue distribution of the complex scaled Hamiltonian
for the $3 \alpha$ system 
obtained by Kurokawa and Kat\={o}~\cite{Kurokawa1}
using the real-range Gaussian basis functions. 
The scaling angle is $\theta=16^\circ$.
The $0^+_3$ state was predicted at  $E_{res}=1.66-i\, 0.74$ \mbox{MeV},
not localized from the  [$\alpha+\alpha+\alpha$] + 
[$^8{\textrm{Be}}(0^+)+\alpha$] continuum 
(see the text). This figure is taken from Ref.~\cite{Kurokawa1}.
}%
\label{fig:fig4-kurokawa}%
\end{figure}

Figure \ref{fig:fig4-kurokawa}, taken from Ref.~\citen{Kurokawa1},
shows the $0^+$ eigenvalue distribution of the complex scaled Hamiltonian
calculated  with the real-range Gaussian basis functions. 
The scaling angle $\theta=16^\circ$ was the maximum angle
available in the calculation.
The new broad $0^+_3$ state was predicted at  
$E_{res}=1.66 -i\, 0.74$ \mbox{MeV}, 
but this complex energy is not isolated from the 
[$\alpha+\alpha+\alpha$]+[$^8{\textrm{Be}}(0^+)+\alpha$]
continuum states in Fig.~\ref{fig:fig4-kurokawa} at $\theta=16^\circ$.
The energy was derived by an extrapolation based on the
ACCC+CSM (see Fig. 2 in Ref.~\citen{Kurokawa1}).

\begin{figure}[!ht]
\begin{minipage}{0.48\textwidth}
\centering
  \includegraphics[width=6.7cm,height=6.7cm,clip]{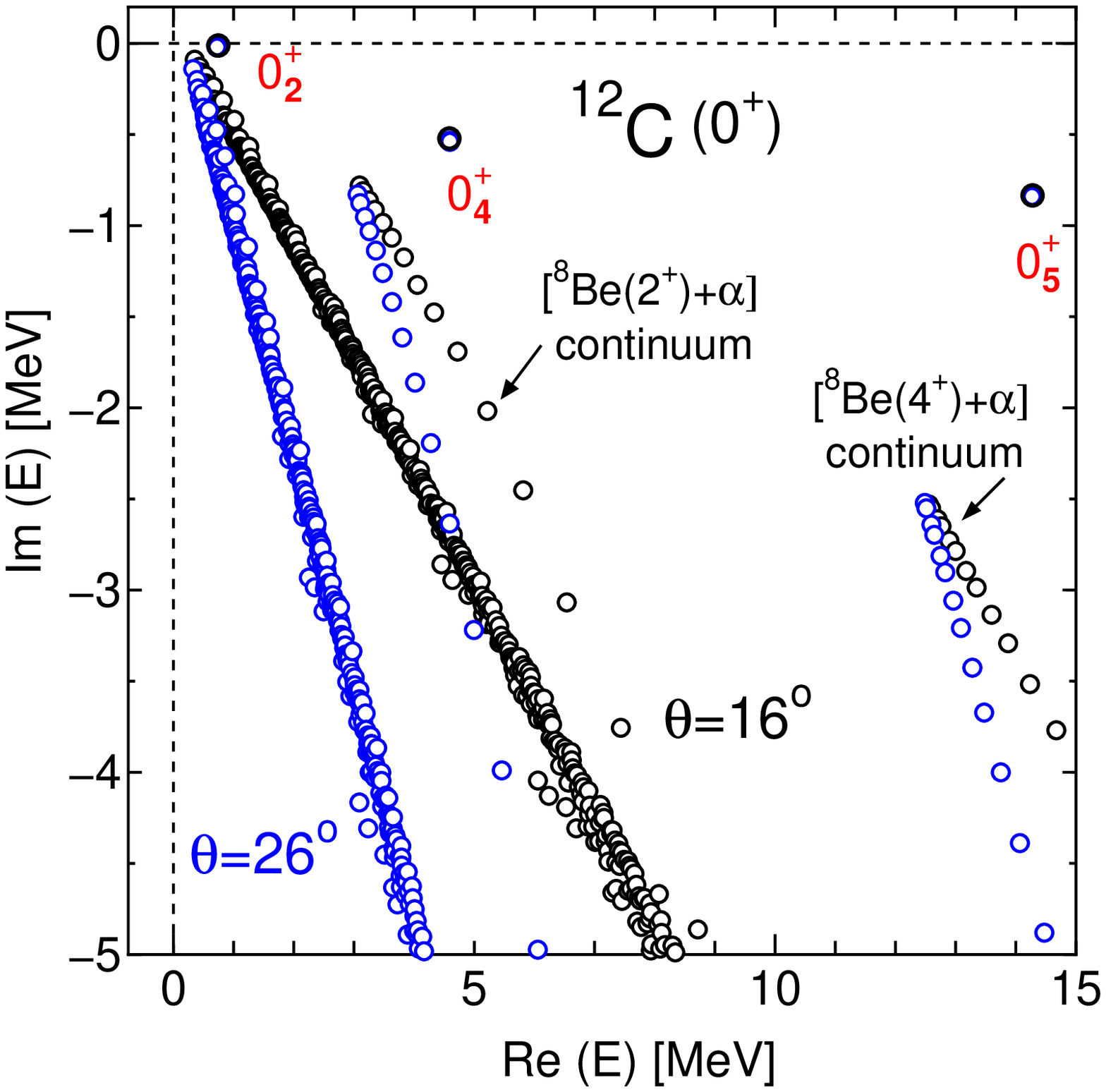}%
\caption{
The $0^+$ eigenvalue distribution of the complex scaled Hamiltonian
for the $3 \alpha$ system with the use of the complex-range Gaussian
basis set in Table~\ref{tab:J0-parameter}.
The scaling angles are $\theta=16^\circ$ (black) and $26^\circ$ (blue).
This figure is to be compared with Fig.~\ref{fig:fig4-kurokawa}.
}%
\label{fig:fig5-J0pole}%
\end{minipage}%
\hspace{\fill}
\begin{minipage}{0.48\textwidth}
\vspace*{-0.1cm}
\centering
  \includegraphics[width=6.7cm,height=6.7cm,,clip]{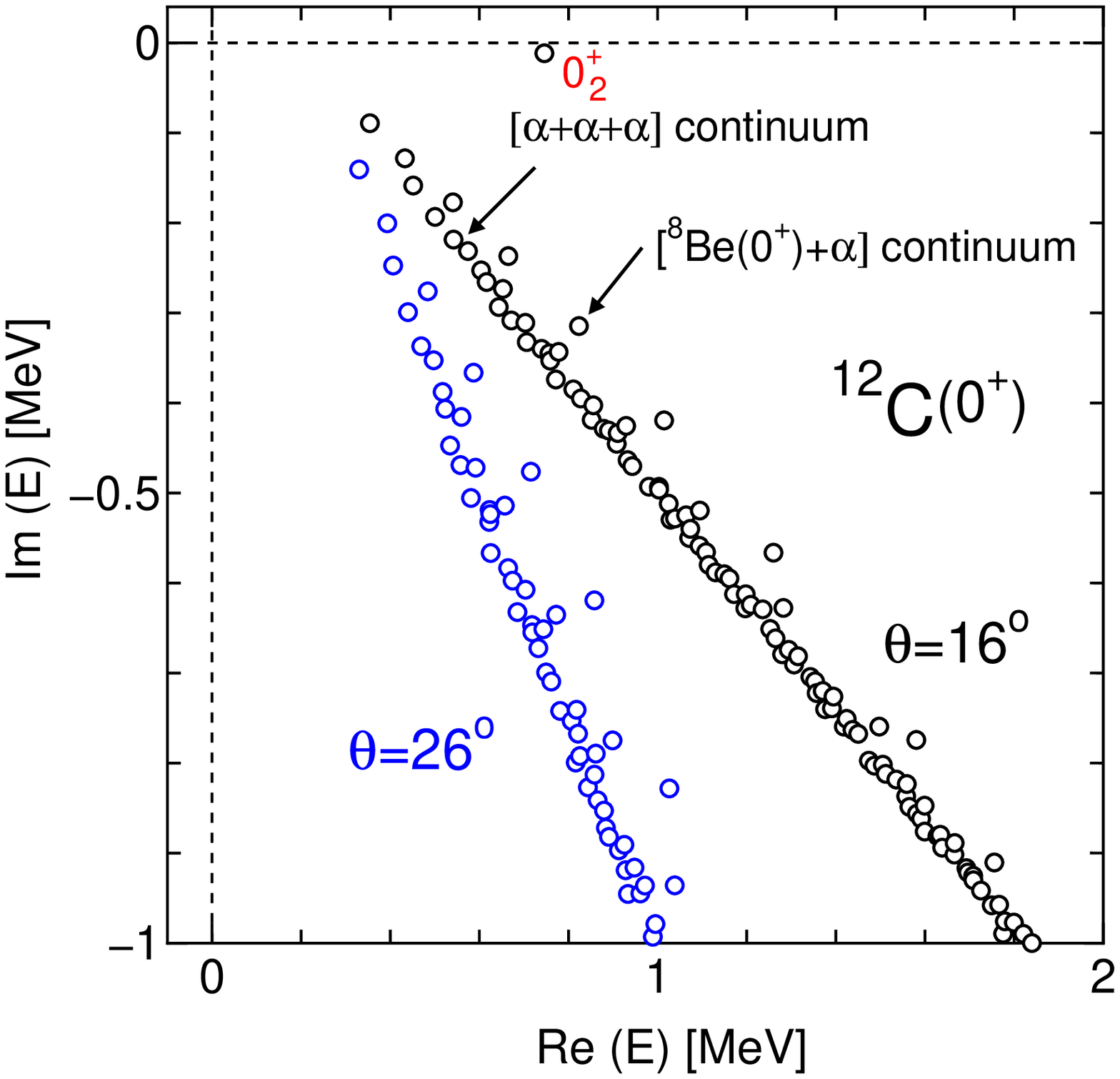}%
\caption{
The low-energy part of Fig.~\ref{fig:fig5-J0pole} is enlarged to show
that the  [$^8{\textrm{Be}}(0^+)+\alpha$] continuum is 
distinguishable  from the [$\alpha+\alpha+\alpha$] one
by $\sim 0.1$ \mbox{MeV} which corresponds to the energy of
$^8{\textrm{Be}}(0^+)$ measured from the $\alpha$-$\alpha$ threshold.
}%
\label{fig:fig6-J0low}%
\end{minipage}
\end{figure}

Figure \ref{fig:fig5-J0pole} illustrates the result of the present 
CSM calculation for the $J=0^+$ states
at $\theta=16^\circ$ and $26^\circ$.
The low-energy part is enlarged in Fig.~6.
All the nonlinear parameters 
used in the calculation are listed 
in Table~\ref{tab:J0-parameter}.   
The parameters for the Gaussian ranges 
are given in  round numbers
but further optimization of them does not significantly improve the 
present result; the same is for the ranges of the $J=2^+$ and $4^+$ states. 
The slightly larger basis set for $l=L=4$ is necessary to precisely
generate  the [$^8{\textrm{Be}}(4^+)+\alpha$] continuum.

The [$\alpha+\alpha+\alpha$] continuum for $\theta=16^\circ$ 
is much less scattered than that in Fig.~\ref{fig:fig4-kurokawa}  and 
the [$^8{\textrm{Be}}(0^+)+\alpha$] continuum is distinguishable in Fig.~6
from the former continuum by $\sim 0.1$ \mbox{MeV} which corresponds to the
energy of the $^8{\textrm{Be}}(0^+)$ resonance measured from the 
$\alpha$-$\alpha$ threshold. 
  It is to be emphasized that, in Fig.~\ref{fig:fig5-J0pole}, 
the localized resonance poles of
the $0^+_2$, $0^+_4$ and $0^+_5$ states for $\theta=26^\circ$
remain at the same places of them
for $\theta=16^\circ$ and that the complex energies of 
the three-body continuum
still form a narrow straight band with little scattered members.
The  energies and widths of those resonances 
are almost the same
as those obtained in Ref.~\citen{Kurokawa1}.

\begin{table}[t]
\caption{ All the nonlinear parameters of the $J=0^+$ 
three-body complex-range
Gaussian basis functions for $^{12}$C
used in the CSM calculation of Fig.~\ref{fig:fig5-J0pole} at $\theta=16^\circ$ and $26^\circ$.
Total number of the basis is $\gamma_{\textrm{max}}= 3200$ 
with $\omega=\pi/2$.
}
\centering
\label{tab:J0-parameter}
\begin{tabular}{cccccccccc}
\noalign{\vskip 0.2 true cm} 
\hline
\noalign{\vskip 0.2 true cm} 
   \multicolumn{4}{c} 
{$J=0^+$ $\qquad $  $r^l e^{-(1 \pm i\omega)(r/r_n)^2}$  } 
 &&  \multicolumn{4}{c} 
{ $R^l e^{-(1 \pm i\omega)(R/R_N)^2}$ } & $\omega=\pi/2$ \\
\noalign{\vskip 0.1 true cm} 
   \multispan4 {\hrulefill} & \qquad &   \multispan4 {\hrulefill} & \\
\noalign{\vskip 0.1 true cm} 
 $l $ & $n_{\textrm{max}} $ & $r_1$ & $r_{n_{\textrm{max}}}$ 
& \qquad & $L$ & $N_{\textrm{max}} $ & $R_1$  & $R_{N_{\textrm{max}}}$ 
& $\;$ number\\
        &              &      [fm] &             [fm]  &
 \qquad &       &      &       [fm] &            [fm] &
 $\;$ of basis\\
\noalign{\vskip 0.1 true cm} 
\hline
\vspace{-3 mm} \\
 0   &  16  &  0.3 & 30.0&  \qquad  & 0 & 16  &  0.5 & 40.0
 &  1024 \\
 2   &  16  &  0.6 & 30.0&  \qquad & 2 & 16  &  1.0 &  40.0 
 &  1024 \\
 4   &  18  &  1.0 & 30.0&  \qquad & 4 & 16  &  1.5 &  40.0 
 &  1152 \\
\vspace{-3 mm} \\
\hline
\end{tabular}
\end{table}

\subsection{The new  $0^+_3$ resonance}

In order to investigate the new $0^+_3$ state that was predicted by
Kurokawa and Kat\={o}~\cite{Kurokawa1,Kurokawa2},
we performed the CSM calculation for
the scaling angles from $\theta=22^\circ$ up to $36^\circ$. 
These large angles are required to reveal explicitly
such a low-lying broad resonance separated from
the [$\alpha+\alpha+\alpha$] + [$^8{\textrm{Be}}(0^+)+\alpha$]
continuum states.
The employed set of the three-body complex-range Gaussian basis functions 
is listed in Table~\ref{tab:J0-3-parameter}. 
Such a larger number of the  basis is 
necessary for this purpose.

\begin{table}[!ht]
\caption{ All the nonlinear parameters of the $J=0^+$ 
three-body complex-range
Gaussian basis functions for $^{12}$C
used in the CSM calculation of Fig.~\ref{fig:fig7-J03pole}
for $\theta=22^\circ \mbox{ to } 36^\circ$
to reveal the $0^+_3$ resonance state.  Total number of basis
is $\gamma_{\textrm{max}}= 4448$ with $\omega=\pi/2$.
}
\centering
\label{tab:J0-3-parameter}
\begin{tabular}{cccccccccc}
\noalign{\vskip 0.2 true cm} 
\hline
\noalign{\vskip 0.2 true cm} 
   \multicolumn{4}{c} 
{$J=0^+$ $\qquad $   $r^l e^{-(1 \pm i\omega)(r/r_n)^2}$  } 
&$\;\;\;$ &  \multicolumn{4}{c} 
{  $R^l e^{-(1 \pm i\omega)(R/R_N)^2}$ } & $\omega=\pi/2$ \\
\noalign{\vskip 0.1 true cm} 
   \multispan4 {\hrulefill} & \qquad &   \multispan4 {\hrulefill} & \\
\noalign{\vskip 0.1 true cm} 
 $l $ & $n_{\textrm{max}} $ & $r_1$ & $r_{n_{\textrm{max}}}$ 
& \qquad & $L$ & $N_{\textrm{max}} $ & $R_1$  & $R_{N_{\textrm{max}}}$ 
 & $\;$ number\\
        &              &      [fm] &             [fm]  &
 \qquad &       &      &       [fm] &            [fm]  & 
$\;$ of basis\\
\noalign{\vskip 0.1 true cm} 
\hline
\vspace{-2 mm} \\
 0   &  22  &  0.3 & 40.0&  \qquad  & 0 & 22  &  0.5 & 40.0
 &  1936 \\
 2   &  22  &  0.6 & 40.0&   \qquad & 2 & 22  &  1.0 &  40.0 
 &  1936 \\
 4   &  12  &  1.0 & 30.0&  \qquad & 4 & 12  &  1.5 &  30.0 
 &   $\;\,$576 \\
\vspace{-3 mm} \\
\hline
\end{tabular}
\end{table}

\begin{figure}[!ht]
\begin{minipage}{0.48\textwidth}
\centering
  \includegraphics[width=6.7cm,height=6.7cm,clip]{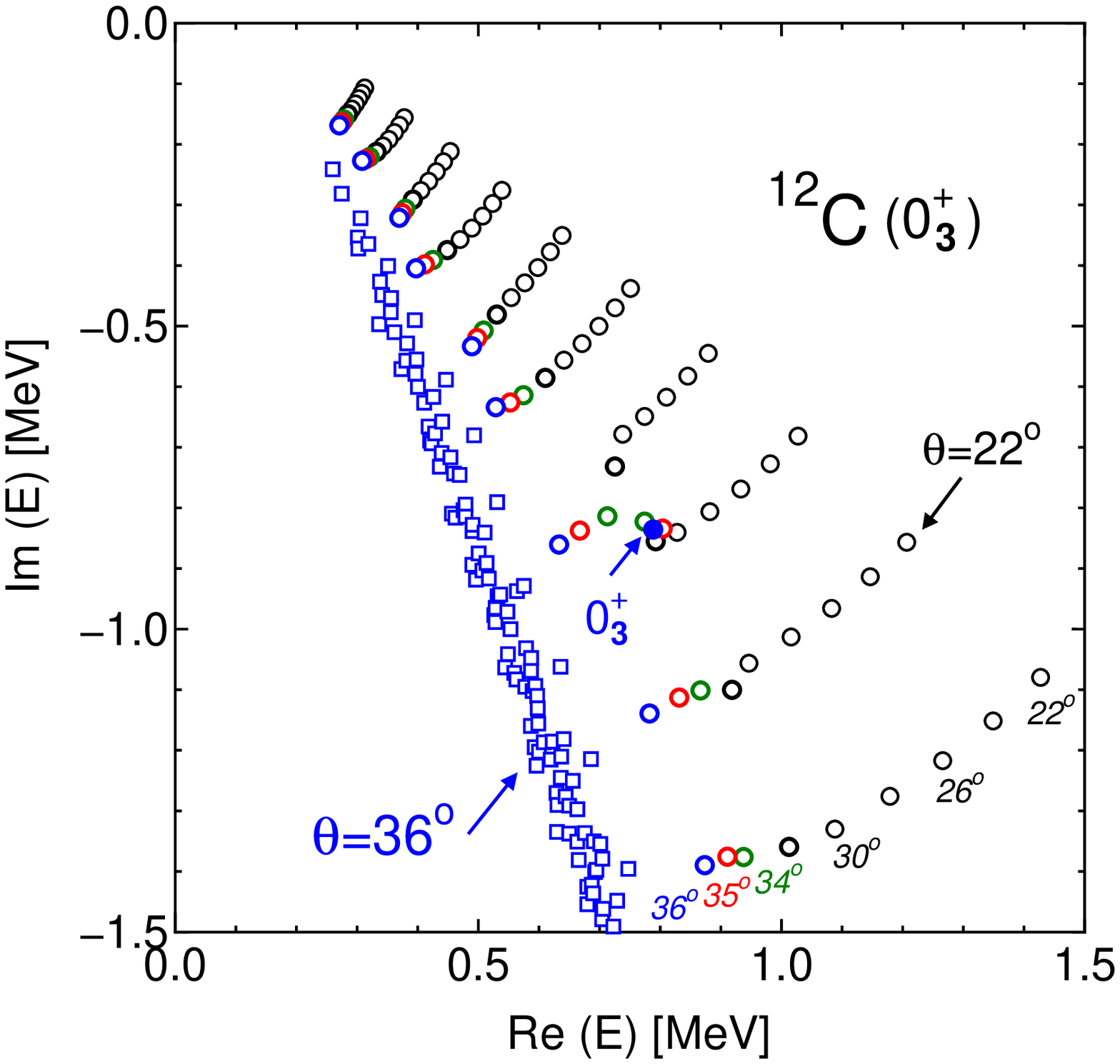}%
\caption{
 The $0^+$ eigenvalue distribution of the complex scaled 
Hamiltonian in which the angle  $\theta$
is varied from $22^\circ$  to $36^\circ$.
The $0^+_3$ resonance appears, 
as the closed blue circle ($36^\circ$), at $E_{res}=0.79 - i\, 0.84$ \mbox{MeV}. 
Only for $\theta=36^\circ$
both the continua of [$\alpha+\alpha+\alpha$] (open blue boxes)
and [$^8{\textrm{Be}} (0^+)+\alpha$] (open blue circles)
are given, but
the former is omitted for \mbox{$\theta < 36^\circ$} for
clarity of the figure.
}%
\label{fig:fig7-J03pole}%
\end{minipage}
\hspace{\fill}
\begin{minipage}{0.48\textwidth}
\vspace*{-0.4cm}
\centering
  \includegraphics[width=6.7cm,height=6.7cm,clip]{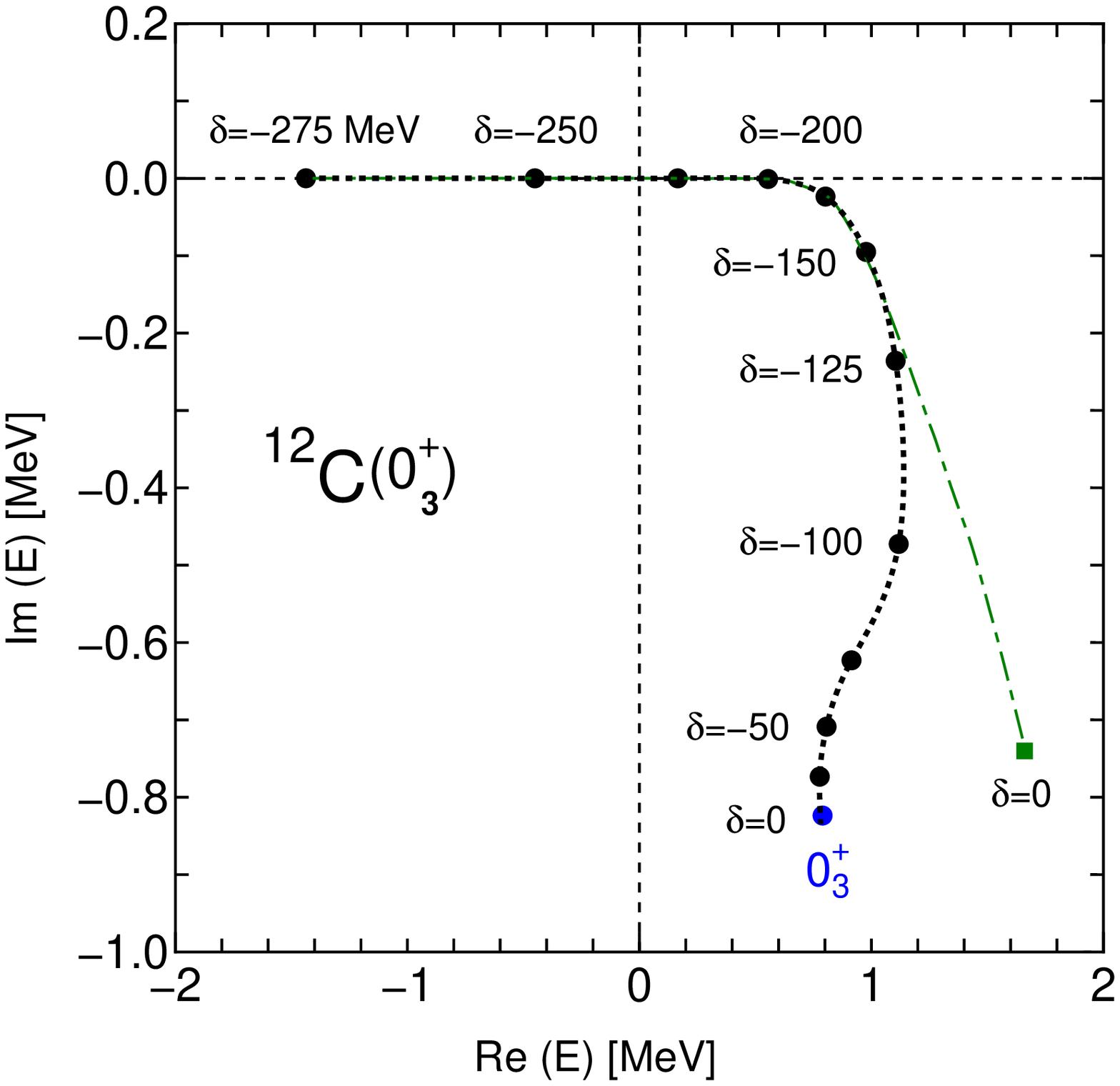}%
\caption{
Trajectory of the $0^+_3$ state obtained by changing the
strength parameter $\delta$ 
of the auxiliary three-body potential (\ref{eq:V-auxil}).
The blue closed circle at $\delta=0$ corresponds to
the $0^+_3$ resonance in Fig.~\ref{fig:fig7-J03pole}. The green box denotes the
$0^+_3$ state predicted by Kurokawa 
and Kat\={o}~\cite{Kurokawa1,Kurokawa2}
on the basis of the extrapolation (the dash-dotted green curve, taken
from Ref.~\citen{Kurokawa1})
using the ACCC+CSM.
}%
\label{fig:fig8-J0traj}%
\end{minipage}
\end{figure}

In Fig.~\ref{fig:fig7-J03pole}, we illustrate the $0^+$ distribution of
complex eigenvalues for $\theta=22^\circ$ up to $36^\circ$.  
Only for $\theta=36^\circ$
both the [$\alpha+\alpha+\alpha$] continuum (open blue boxes)
and the [$^8{\textrm{Be}} (0^+)+\alpha$] continuum (open blue circles)
are given, but
the former is omitted for \mbox{$\theta < 36^\circ$} to avoid
complexity of the figure.
We observe a converged resonance pole at $E_{res}=0.79 - i\,0.84$ \mbox{MeV}
(the closed blue circle) and 
identify it as the third
$0^+$ state that was predicted in Ref.~\citen{Kurokawa1}.   
The position and width of the resonance, however, differ slightly 
from the result in Ref.~\citen{Kurokawa1}, $E_{res}=1.66 -i\,0.74$ \mbox{MeV}.

Reason of this difference is explained with Fig.~\ref{fig:fig8-J0traj} which 
is to be compared with Fig.~2 in Ref.~\citen{Kurokawa1}.
Figure \ref{fig:fig8-J0traj} illustrates the 
trajectory of the $0^+_3$ state on the complex energy plane,
that was obtained by changing the
strength parameter $\delta$ 
of the auxiliary three-body potential, Eq.~(4) in Ref.~\citen{Kurokawa1},
added to the Hamiltonian (\ref{eq:Hamiltonian}),
\begin{equation}
\label{eq:V-auxil}
  V_{\textrm{aux.}}= \delta \,
    \exp \left[\, -\mu ( r_1^2 + r_2^2 + r_3^2) \,\right],
  \qquad \mu=0.15 \:\textrm{fm}^{-2}.
\end{equation}
The closed blue circle for $\delta=0$ in Fig.~\ref{fig:fig8-J0traj}
is the same as that for the $0^+_3$ 
resonance in Fig.~\ref{fig:fig7-J03pole}.
 On the other hand, in Ref.~\citen{Kurokawa1},
the direct CSM calculation of the $0^+_3$ resonance 
was not possible when the auxiliary $3 \alpha$
potential is less attractive than $\delta= -120$ \mbox{MeV}.
 The green box that indicates the
$0^+_3$ state in Ref.~\citen{Kurokawa1} was therefore estimated
by the extrapolation (the dash-dotted green curve)
using the ACCC+CSM.
We thus understand that the difference in  
the resonance-pole position between the two calculations comes from 
the error of the extrapolation.

We conclude that
we have confirmed the prediction by Kurokawa and Kat\={o}~\cite{Kurokawa1} 
about the appearance of a new $0^+_3$ broad resonance 
that is located slightly above the Hoyle state $(0^+_2)$.
As long as the structure of
the $0^+_3$ state is concerned, it is interesting to see that,
in Fig.~\ref{fig:fig7-J03pole}, the converged pole of the state 
is generated from the [$^8{\textrm{Be}} (0^+)+\alpha$] continuum 
during the scaling angle is rotated up to $\theta=36^\circ$.
Therefore,  the $0^+_3$ state is considered to be dominantly composed of
the [$^8{\textrm{Be}} (0^+)+\alpha$] configuration.
Kurokawa and Kat\={o}~\cite{Kurokawa2} pointed out that
the $0^+_3$ state has a similar property to the $0^+_2$ state 
and the former may be a higher nodal state of the latter having
the [$^8{\textrm{Be}} (0^+) + 0^+(\alpha)]$ configuration mainly,
which is consistent with our conjecture.

\begin{table}[!ht]
\caption{ All the nonlinear parameters of the $J=2^+$ 
three-body complex-range
Gaussian basis functions for $^{12}$C
used in the CSM calculation of Fig.~\ref{fig:fig9-J2pole} at $\theta=16^\circ$ and $26^\circ$.
Total number of the basis is $\gamma_{\textrm{max}}= 6400$ 
with $\omega=\pi/2$. 
}
\centering
\label{tab:J2-parameter}
\begin{tabular}{cccccccccc}
\noalign{\vskip 0.2 true cm} 
\hline
\noalign{\vskip 0.2 true cm} 
   \multicolumn{4}{c} 
{$J=2^+$ $\qquad $   $r^l e^{-(1 \pm i\omega)(r/r_n)^2}$  } 
&$\;\;\;$ &  \multicolumn{4}{c} 
{  $R^l e^{-(1 \pm i\omega)(R/R_N)^2}$ } & $\omega=\pi/2$ \\
\noalign{\vskip 0.1 true cm} 
   \multispan4 {\hrulefill} & \qquad &   \multispan4 {\hrulefill} & \\
\noalign{\vskip 0.1 true cm} 
 $l $ & $n_{\textrm{max}} $ & $r_1$ & $r_{n_{\textrm{max}}}$ 
& \qquad & $L$ & $N_{\textrm{max}} $ & $R_1$  & $R_{N_{\textrm{max}}}$ 
& $\;$ number\\
         &              &       [fm] &            [fm]  
& \qquad &       &      &       [fm] &            [fm] 
& $\;$ of basis\\
\noalign{\vskip 0.1 true cm} 
\hline
\vspace{-3 mm} \\
 0   &  16  &  0.3 & 30.0&  \qquad  & 2 & 16  &  1.0 & 40.0
  & 1024 \\
 2   &  16  &  0.6 & 30.0&   \qquad & 0 & 16  &  0.5 & 40.0 
  & 1024 \\
 2   &  16  &  0.6 & 30.0&   \qquad & 2 & 16  &  1.0 & 40.0 
  & 1024 \\
 2   &  16  &  0.6 & 30.0&  \qquad & 4 & 16 &  1.5 &  40.0 
  & 1024 \\
 4   &  18  &  1.0 & 30.0&  \qquad & 2 & 16 &  1.0 &  40.0 
 & 1152 \\
 4   &  18  &  1.0 & 30.0&  \qquad & 4 & 16  &  1.5 &  40.0 
  & 1152 \\
\vspace{-3 mm} \\
\hline
\end{tabular}
\end{table}

\begin{figure}[!ht]
\begin{minipage}{0.48\textwidth}
\centering
  \includegraphics[width=6.7cm,height=6.7cm,clip]{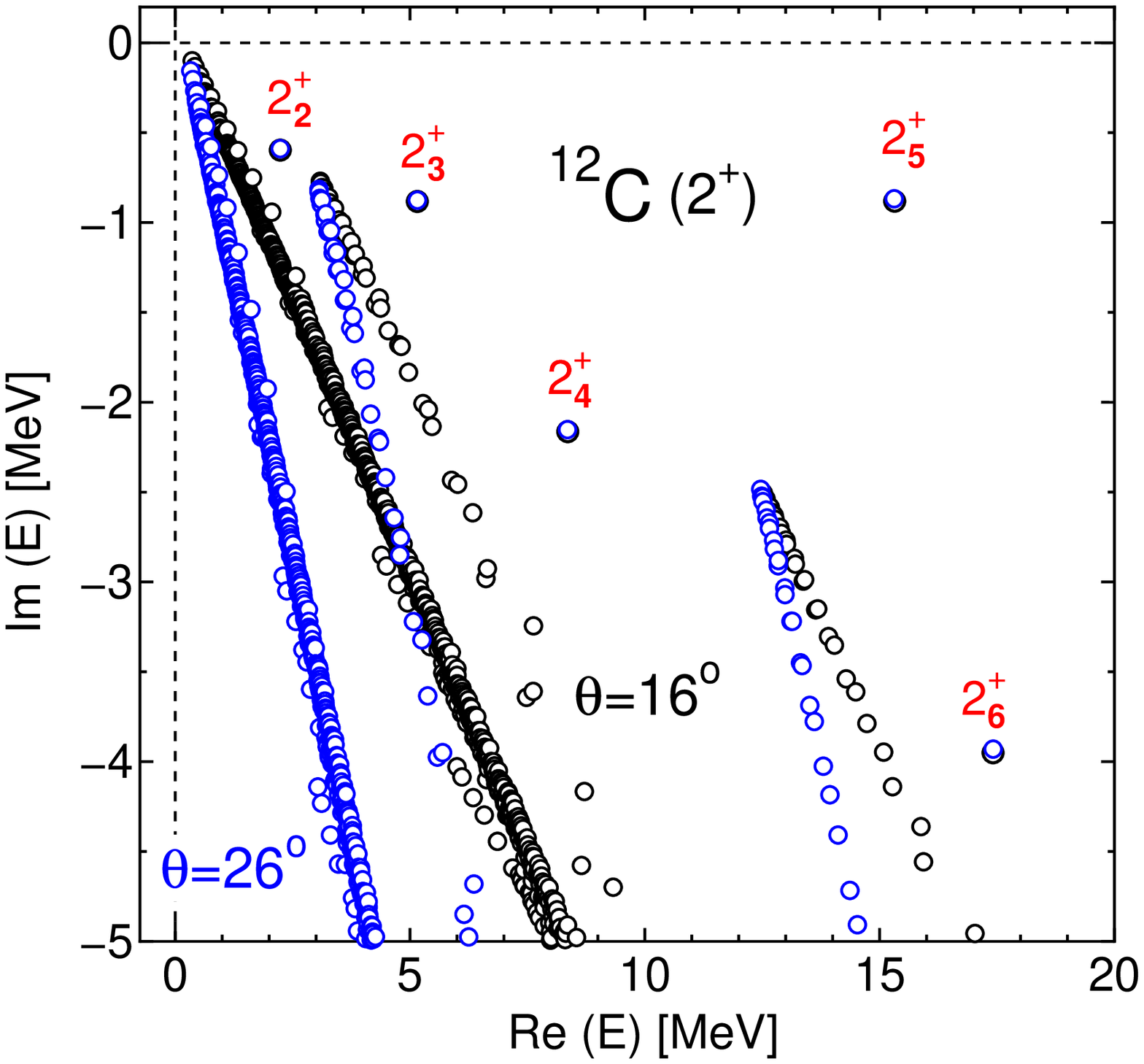}%
\caption{
The $2^+$ eigenvalue distribution of the complex scaled Hamiltonian
calculated with the
complex-range Gaussian basis set in Table~\ref{tab:J2-parameter}.
The scaling angles are $\theta=16^\circ$ (black) and $26^\circ$ (blue).
This figure is to be compared with Fig.~5 in Ref.~\citen{Kurokawa2}
at $\theta=16^\circ$.
}%
\label{fig:fig9-J2pole}%
\end{minipage}
%
\hspace{0.3cm}
\begin{minipage}{0.48\textwidth}
\centering
  \includegraphics[width=6.7cm,height=6.7cm,clip]{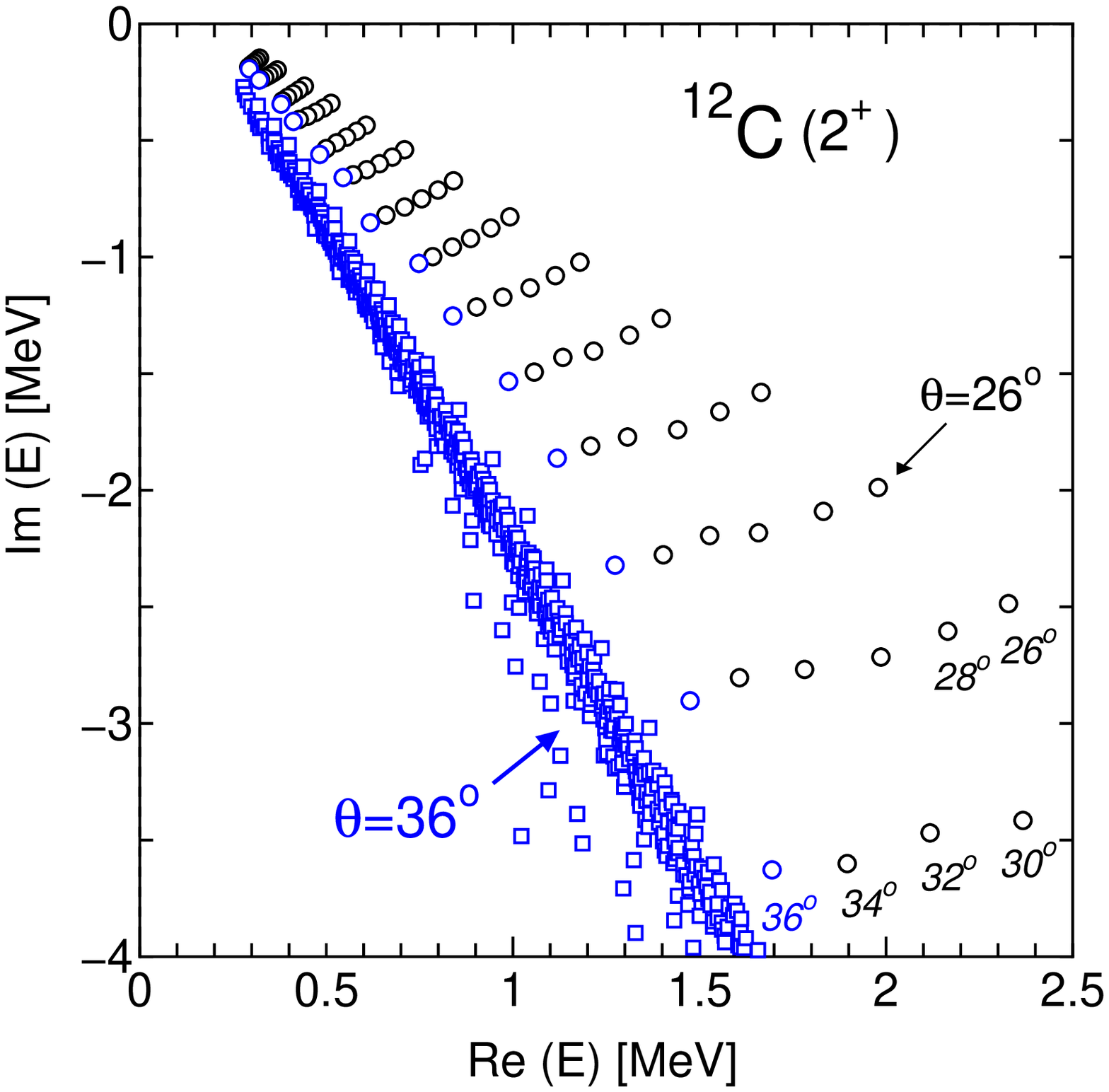}%
\caption{
Low-energy part of the $2^+$ eigenvalue 
distribution for  $\theta=26^\circ$  to $36^\circ$.
Any $2^+$ resonance,
like the $0^+_3$ state in Fig.~\ref{fig:fig7-J03pole}, does not appear
in the upper-right side of 
the [$\alpha+\alpha+\alpha$] continuum (blue boxes)  at 
$\theta=36^\circ$.  See also the caption of Fig.~\ref{fig:fig7-J03pole}.
}%
\label{fig:fig10-J2low}%
\end{minipage}
\end{figure}

\subsection{The $2^+$ resonances}

Figure \ref{fig:fig9-J2pole} illustrates 
the calculated $2^+$ eigenvalue distribution 
of the complex scaled Hamiltonian for the $3 \alpha$ system. 
The scaling angles are $\theta=16^\circ$ and $26^\circ$.
This figure is much more precise than Fig.~\ref{fig:fig5-J0pole}
($\theta=16^\circ$)
in Ref.~\citen{Kurokawa2} for the $2^+$ eigenvalue distribution.
All the nonlinear parameters
used for calculating Fig.~\ref{fig:fig9-J2pole} are listed in Table~\ref{tab:J2-parameter}.
Total number of the basis is $\gamma_{\textrm{max}}= 6400$ 
with $\omega=\pi/2$.

Calculated five resonances denoted as $2^+_2, \ldots, 2^+_6$
appear at  almost the same complex energies of those obtained in
Ref.~\citen{Kurokawa2}.
We observe no other $2^+$ resonance at low  energies.
As shown in Fig.~\ref{fig:fig10-J2low}, even if the scaling angle is
increased up to $\theta=36^\circ$, any new $2^+$ resonance,
like the $0^+_3$ resonance in Fig.~\ref{fig:fig7-J03pole}, does not appear
in the upper-right side of 
the [$\alpha+\alpha+\alpha$] continumm  at 
$\theta=36^\circ$.

\begin{figure}[b]
\centering
  \includegraphics[width=6.7cm,height=6.7cm,clip]{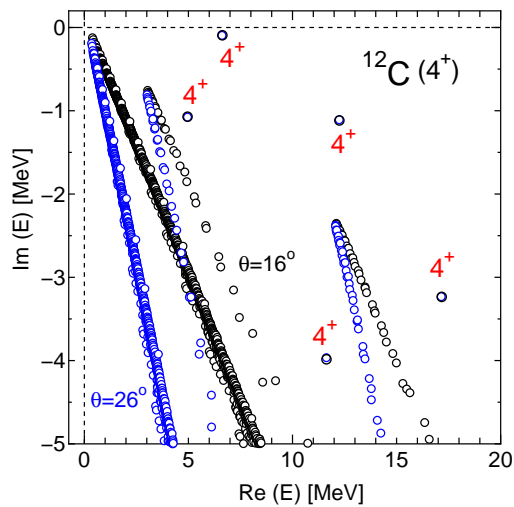}%
\caption{
The $4^+$ eigenvalue distribution of the complex scaled Hamiltonian
for the $3 \alpha$ system
calculated with the complex-range Gaussian basis set in Table~\ref{tab:J4-parameter}.
The scaling angles are $\theta=16^\circ$ (black) and $26^\circ$ (blue).
The lowest resonance at $E_{res}=4.96 -i\,1.1$ MeV 
does not corresponds to the observed $4^+_1$ state at $E_r=6.81$ MeV
which is reproduced by the second $4^+$ resonance in this figure.
See the text about this problem.
}%
\label{fig:fig11-J4pole}%
\end{figure}

\subsection{The $4^+$ resonances}

The calculated $4^+$ eigenvalue distribution of the complex
scaled Hamiltonian is illustrated in Fig.~\ref{fig:fig11-J4pole}. 
The resonance parameters are summarized in Table~\ref{tab:J4-table}
 together with
the result by Kurokawa and Kat\={o}~\cite{Kurokawa2}.
All the nonlinear parameters of the basis set are listed
 in Table~\ref{tab:J4-parameter}.
We note that the lowest $4^+$ resonance 
at $E_{res}=4.96 - i\, 1.1$ MeV in Fig.~\ref{fig:fig11-J4pole} is missing
in Ref.~\citen{Kurokawa2}, where the lowest one 
is given at $E_{res}=6.82 - i \,0.12$ MeV,
but it corresponds to the second $4^+$ state~in~Fig.~\ref{fig:fig11-J4pole}.

This situation causes a serious problem 
in the determination    
of the $3\alpha$ potential $V_{3\alpha}$
of Eq.~(\ref{eq:V-alpha-alpha}). The strongly repulsive factor of 
$V_{3\alpha}^{4^+}=150.0$ MeV was so chosen in Ref.~\citen{Kurokawa2}
that the calculated lowest $4^+$ state can reproduce the
observed value of  $E_{res}(4^+_1)=6.808 - i\,0.258$ MeV.
However, by the introduction of the 
repulsive $3\alpha$ potential,
the original lowest $4^+$ state at $-1.66$~MeV 
(with $V_{3\alpha}^{4^+}=0$) is pushed up to 
$E_{res}=6.61 - i \,0.10$ MeV which corresponds to 
the observed $4^+_1$ state,
but the second $4^+$ state (with $V_{3\alpha}^{4^+}=0$)
remains almost unaffected at $E_{res}(4^+)= 4.96 - i\,1.1$ MeV 
that becomes the lowest $4^+$ state in Fig.~\ref{fig:fig11-J4pole}.

 Therefore, we understand that 
the introduction of the $3\alpha$ potential $V_{3\alpha}$
does not work for the  $J=4^+$ states
even if the strength is given dependently on $J$.
Any appropriate determination of the interaction 
in the $3\alpha$ system will be required in future
$3\alpha$ OCM-CSM calculations.

\begin{table}[t]
\caption{  
Calculated resonance parameters for the $J=4^+$ states in $^{12}$C 
by the present work  and
Ref.~\citen{Kurokawa2} which employ the same interaction.
All quantities are given in MeV.
}
\vspace{5pt}
\centering
\label{tab:J4-table}
\begin{tabular}{c c r@{.}l r@{.}l r@{.}l c r@{.}l r@{.}l r@{.}l }
\hline  \\ [-8pt]
  $^{12}$C    & \phantom{??}
              & \multicolumn{6}{c}{present work}
              & \phantom{??}
              & \multicolumn{6}{c}{Ref.~\citen{Kurokawa2}}   \\ [4pt]
\cline{3-8} \cline{10-15}  \\ [-8pt]
  $J^{\pi}$   &{}
              & \multicolumn{2}{c}{$E_{x}$}
              & \multicolumn{2}{c}{$E_{r}$}
              & \multicolumn{2}{c}{$\varGamma$} &{}
              & \multicolumn{2}{c}{$E_{x}$}
              & \multicolumn{2}{c}{$E_{r}$}
              & \multicolumn{2}{c}{$\varGamma$}    \\ [2pt]
\hline  \\ [-8pt]
  $4^{+}$  &{}&  12&25  &  4&96  &   2&2  &{}
              &  \multicolumn{2}{c}{---}
              &  \multicolumn{2}{c}{---}
              &  \multicolumn{2}{c}{---}         \\ [1pt]
  $4^{+}$  &{}&  13&91  &  6&61  &   0&20  &{}
              &  14&11  &  6&82  &   0&24        \\ [1pt]
  $4^{+}$  &{}&  18&92  & 11&62  &   8&0   &{}
              &  \multicolumn{2}{c}{---}
              &  \multicolumn{2}{c}{---}
              &  \multicolumn{2}{c}{---}         \\ [1pt]
  $4^{+}$  &{}&  19&53  & 12&23  &   2&2   &{}
              &  20&39  & 13&1   &   3&4         \\ [1pt]
  $4^{+}$  &{}&  24&41  & 17&11  &   6&3   &{}
              &  \multicolumn{2}{c}{---}
              &  \multicolumn{2}{c}{---}
              &  \multicolumn{2}{c}{---}         \\ [2pt]
\hline
\end{tabular}
\end{table}

\begin{table}[t]
\caption{ All the nonlinear parameters of the $J=4^+$ 
three-body complex-range
Gaussian basis functions for $^{12}$C
used in the CSM calculation of Fig.~\ref{fig:fig11-J4pole} at $\theta=16^\circ$ and $26^\circ$.
Total number of the basis is $\gamma_{\textrm{max}}= 8640$ 
with $\omega=\pi/2$. 
}
\centering
\label{tab:J4-parameter}
\begin{tabular}{cccccccccc}
\noalign{\vskip 0.2 true cm} 
\hline
\noalign{\vskip 0.2 true cm} 
   \multicolumn{4}{c} 
{$J=4^+$ $\qquad $   $r^l e^{-(1 \pm i\omega)(r/r_n)^2}$  } 
&$\;\;\;$ &  \multicolumn{4}{c} 
{  $R^l e^{-(1 \pm i\omega)(R/R_N)^2}$ } & $\omega=\pi/2$ \\
\noalign{\vskip 0.1 true cm} 
   \multispan4 {\hrulefill} & \qquad &   \multispan4 {\hrulefill} & \\
\noalign{\vskip 0.1 true cm} 
 $l $ & $n_{\textrm{max}} $ & $r_1$ & $r_{n_{\textrm{max}}}$ 
& \qquad & $L$ & $N_{\textrm{max}} $ & $R_1$  & $R_{N_{\textrm{max}}}$ 
& $\;$ number\\
        &              &      [fm] &             [fm]  
& \qquad &       &      &       [fm] &            [fm] 
& $\;$ of basis\\
\noalign{\vskip 0.1 true cm} 
\hline
\vspace{-3 mm} \\
 0   &  20  &  0.3 & 30.0&  \qquad  & 4 & 18  &  1.0 & 40.0
  & 1440 \\
 2   &  20  &  0.6 & 30.0&   \qquad & 2 & 18  &  0.5 &  40.0 
  & 1440 \\
 2   &  20  &  0.6 & 30.0&   \qquad & 4 & 18  &  1.0 &  40.0 
  & 1440 \\
 4   &  20  &  0.6 & 30.0&  \qquad & 0 & 18 &  1.5 &  40.0 
  & 1440 \\
 4   &  20  &  1.0 & 30.0&  \qquad & 2 & 18 &  1.0 &  40.0 
 & 1440 \\
 4   &  20  &  1.0 & 30.0&  \qquad & 4 & 18  &  1.5 &  40.0 
  & 1440 \\
\vspace{-3 mm} \\
\hline
\end{tabular}
\end{table}

\section{Summary}
\label{sect:Summary}

The authors have proposed  
to use the complex-range Gaussian basis functions,
\{$r^l\:e^{- ( 1 \pm i\, \omega )(r/r_n)^2 }Y_{lm}({\widehat{\mathbf{r}}})$;
 $r_n$ in a geometric progression\},
in the CSM calculations of three-body resonances 
in place of the real-range Gaussians 
that are often employed in the literature.  
The former-type Gaussians are very suitable for describing
short-range correlations, long-range asymptotic decaying amplitudes
and highly oscillating behavior in few-body systems
as well as they are tractable in calculating 
the Hamiltonian matrix elements
with transformation between different sets of Jacobi 
coordinates~\cite{Hiyama03}.
Therefore, they are particularly useful in the CSM calculations
when representing the resonant and nonresonant
continuum states that become quite  oscillatory 
as the scaling angle $\theta$ increases; this enables us, in the
study of broad three-body resonances, to take 
much larger angles than those considered before
and to have a possibility of  observing new broad resonance poles.

The present method has been applied to 
the $3\alpha$ resonances in $^{12}$C with $J=0^+, 2^+$ and $4^+$.
The result was compared with that obtained 
by Kurokawa and Kat\={o}~\cite{Kurokawa1,Kurokawa2} where the real-range 
Gaussians were employed to expand the $3\alpha$ wave function.
In Table~\ref{tab:Summary-energy}, we summarize the calculated energies 
and widths 
of the states with $J=0^+$ and $2^+$ together with the result by 
Refs.~\citen{Kurokawa1} and \citen{Kurokawa2} and the experimental data.
The result for the $4^+$ resonances, having a problem 
in the interaction employed, was summarized in Table~\ref{tab:J4-table}.

\begin{table}[!t]
\vspace*{-3mm}
\caption{  
Summary of the calculated result for the $J=0^+$ and $2^+$ in $^{12}$C 
by the present work  together with the
result by Refs.~\citen{Kurokawa1} and \citen{Kurokawa2}
and the experimental data. The data for $E_x=9.04$, $9.84$
and 10.56 \mbox{MeV} 
are taken from Ref.~\citen{Itoh2011};
see the footnote in Section \ref{sect:Introduction}.
The other data are taken from Ref.~\citen{Ajzenberg}.
The model and interaction are the same between the two calculations.
All quantities are given in MeV.
}
\vspace{5pt}
\centering
\label{tab:Summary-energy}
\begin{tabular}{c  r@{.}l r@{.}l r@{.}l  r@{.}l r@{.}l r@{.}l
                   r@{.}l r@{.}l r@{.}l  }
\hline  \\ [-6pt]
  $^{12}$C    & \multicolumn{6}{c}{present work}
              & \multicolumn{6}{c}{Refs.~\citen{Kurokawa1} and \citen{Kurokawa2}}
              & \multicolumn{6}{c}{Experimental data}   \\ [6pt]
\cline{2-8} \cline{9-14} \cline{15-19}  \\ [-8pt]
  $J^{\pi}$   & \multicolumn{2}{c}{$E_{x}$}
              & \multicolumn{2}{c}{$E_{r}$}
              & \multicolumn{2}{c}{$\varGamma$}
              & \multicolumn{2}{c}{$E_{x}$}
              & \multicolumn{2}{c}{$E_{r}$}
              & \multicolumn{2}{c}{$\varGamma$}
              & \multicolumn{2}{c}{$E_{x}$}
              & \multicolumn{2}{c}{$E_{r}$}
              & \multicolumn{2}{c}{$\varGamma$}    \\ [3pt]
\hline  \\ [-8pt]
  $0_{1}^{+}$ &   0&00     & $-$7&30    &  \multicolumn{2}{c}{---}
              &   0&00     & $-$7&29    &  \multicolumn{2}{c}{---}
              &   0&00000  & $-$7&2747  &  \multicolumn{2}{c}{---}       \\ [1pt]
  $2_{1}^{+}$ &   4&32     & $-$2&98    &  \multicolumn{2}{c}{---}
              &   4&31     & $-$2&98    &  \multicolumn{2}{c}{---}
              &   4&43891  & $-$2&8358  &  \multicolumn{2}{c}{---}       \\ [1pt]
  $0_{2}^{+}$ &   8&05     &    0&75    &  0&0088
              &   8&05     &    0&76    &  0&0024
              &   7&65420  &    0&3795  &  8&5 $\times 10^{-6}$          \\ [1pt]
  $0_{3}^{+}$ &   8&09     &    0&79    &  1&68
              &   8&95     &    1&66    &  1&48
              &   9&04(9)  &    1&77    &  1&45(18)                      \\ [1pt]
  $2_{2}^{+}$ &   9&54     &    2&24    &  1&2
              &   9&57     &    2&28    &  1&1
              &   9&84(6)  &    2&57    &  1&01(15)                      \\ [1pt]
  $0_{4}^{+}$ &  11&89     &    4&59    &  1&0
              &  11&87     &    4&58    &  1&1
              &  10&56(6)  &    3&29    &  1&42(8)                       \\ [1pt]
  $2_{3}^{+}$ &  12&47     &    5&15    &  1&8
              &  12&43     &    5&14    &  1&9
              &  11&16(5)  &    3&89    &  0&43(8)                       \\ [1pt]
  $2_{4}^{+}$ &  15&67     &    8&36    &  4&3
              &  15&93     &    8&64    &  3&9
              &  15&44(4)  &    8&17    &  1&5(2)                        \\ [1pt]
  $0_{5}^{+}$ &  21&60     &   14&3     &  1&7
              &  21&59     &   14&3     &  1&5
              &  \multicolumn{2}{c}{---}   &  \multicolumn{2}{c}{---}
              &  \multicolumn{2}{c}{---}                                    \\ [1pt]
  $2_{5}^{+}$ &  22&70     &   15&3     &  1&8
              &  22&39     &   15&1     &  1&2
              &  \multicolumn{2}{c}{---}   &  \multicolumn{2}{c}{---}
              &  \multicolumn{2}{c}{---}                                    \\ [1pt]
  $2_{6}^{+}$ &  24&70     &   17&4     &  8&0
              &  24&89     &   17&6     &  6&0
              &  \multicolumn{2}{c}{---}   &  \multicolumn{2}{c}{---}
              &  \multicolumn{2}{c}{---}                                    \\ [3pt]
\hline
\end{tabular}
\end{table}

The distribution of eigenvalues of the complex scaled Hamiltonian
at $\theta=16^\circ$~and $26^\circ$ were found to become 
very  precise (Figs.~\ref{fig:fig5-J0pole}, \ref{fig:fig6-J0low},
~\ref{fig:fig9-J2pole} and \ref{fig:fig11-J4pole}).
The prediction of the new broad $0^+_3$ resonance 
by Kurokawa and Kat\={o}~\cite{Kurokawa1}
was confirmed by our calculation; namely,
as the scaling angle $\theta$ increases up to $36^\circ$,
the $0^+_3$ resonance pole becomes clearly separated from the 
 [$\alpha+\alpha+\alpha$] and [$^8{\textrm{Be}} (0^+)+\alpha$] 
continuum states (Fig.~\ref{fig:fig7-J03pole}).
The slight deviation of the $0^+_3$ resonance energy $E_r$ 
by 0.9~\mbox{MeV} in Ref.~\citen{Kurokawa1}
from our result   
is attributed to the error of the extrapolation~\cite{Kurokawa1} 
of the resonance position by the ACCC+CSM (Fig.~\ref{fig:fig8-J0traj}).
As for the $0^+_4, 0^+_5, 2^+_2, 2^+_3, 2^+_4, 2^+_5$ and $2^+_6$
resonances, we obtained almost the same energies
and widths as those 
in Refs.~\citen{Kurokawa1} and \citen{Kurokawa2}. We did not observe
any  $2^+$ low-lying broad resonance,  like the $0^+_3$,
as long as we increased the scaling angle $\theta$ 
up to $36^\circ$ (Fig.~\ref{fig:fig10-J2low}).

We employed the same interaction for the $3\alpha$ system
as used in  Refs.~\citen{Kurokawa1} and \citen{Kurokawa2}, but 
the calculation did not  satisfactorily well reproduce 
the observed energy of  the important 
Hoyle state $(0^+_2)$, $E_r=0.38$ \mbox{MeV},
with a deviation of some 0.4 \mbox{MeV} higher.  
Furthermore, the strongly repulsive $3\alpha$ potential 
for the $J=4^+$ states, introduced in Ref.~\citen{Kurokawa2}
to reproduce the observed energy of the $4^+_1$ state, 
is found to be not appropriate because 
the lowest $4^+$ state in the present work
is the broad resonance 
at $E_{res}=4.96 -i\, 1.1$ MeV (Fig.~\ref{fig:fig11-J4pole}) which was missing 
in Ref.~\citen{Kurokawa2} and the second $4^+$ state
corresponds to the observed $4^+_1$ state.

We explicitly listed, in small tables, all the nonlinear parameters of 
the basis functions used in the present calculation of 
$^{12}$C $(0^+, 2^+$, $4^+$); 
our method is so transparent. 
 For the comprehensive understanding of 
the $3\alpha$ cluster structure of 
$^{12}$C, however,  use of more improved interactions
is highly desirable in future studies.

\section*{Acknowledgements}

The authors would like to thank Professor K. Kat\={o}
for valuable discussion on the complex scaling method and the
$3\alpha$-cluster structure of $^{12}$C.
Thanks are also due to Dr. T. Matsumoto for helpful discussions
on the numerical complex scaling calculations.
 The authors thank the Yukawa Institute for Theoretical Physics 
at Kyoto University. Discussions during 
the YITP workshop YITP-W-12-19 on ``Resonances and 
non-Hermitian systems in quantum mechanics'' were 
useful to complete this work. 
The numerical calculations were performed
on the HITACHI SR16000 at KEK, 
at Research Institute for Information Technology, Kyushu University 
and at Yukawa Institute for Theoretical Physics, Kyoto University.



\begin{thebibliography}{99}

\bibitem{CSM-ref1}     J.~Aguilar and J.M.~Combes,
      Commun. Math. Phys. \textbf{ 22}, 269 (1971).
\bibitem{CSM-ref2}     E.~Balslev and J.M.~Combes,
      Commun. Math. Phys. \textbf{ 22}, 280 (1971).
\bibitem{CSM-ref3}     B.~Simon,
      Commun. Math. Phys. \textbf{ 27}, 1 (1972).
\bibitem{Ho} Y.K. Ho, Phys. Rep. \textbf{ 99}, 1 (1983).
\bibitem{Moiseyev} N.~Moiseyev, Phys. Rep. \textbf{302}, 212 (1998). 

\bibitem{CSM-review} S. Aoyama, T. Myo, K. Kat\={o}, and K. Ikeda,
Prog. Theor. Phys. \textbf{116}, 1 (2006).

\bibitem{Kamimura88} M.~Kamimura, Phys. Rev. A \textbf{38}, 621 (1988).

\bibitem{Kameyama89}
H. Kameyama, M. Kamimura, and Y. Fukushima, Phys.
Rev. C \textbf{40}, 974 (1989).

\bibitem{Hiyama03} E.~Hiyama, Y.~Kino, and M.~Kamimura, 
Prog. Part. Nucl. Phys. \textbf{51}, 223 (2003).

\bibitem{Hiyama2012ptep} E.~Hiyama, Prog. Theor. Exp. Phys.
(2012) 01A204. 

\bibitem{Kurokawa1} C. Kurokawa and K. Kat\={o}, Phys. Rev. C \textbf{71},
021301 (R) (2005).

\bibitem{Kurokawa2} C. Kurokawa and K. Kat\={o}, Nucl. Phys. A 
\textbf{792}, 82 (2007).

\bibitem{Itoh2011} M.~Itoh, H.~Akimune, M.~Fujiwara, U.~Garg, 
N.~Hashimoto, T.~Kawabata, K.~Kawase, S.~Kishi, T.~Murakami, 
K.~Nakanishi, Y.~Nakatsugawa,  B.~K.~Nayak, S.~Okumura, 
H.~Sakaguchi, H.~Takeda, S.~Terashima, M.~Uchida,
Y.~Yasuda, M.~Yosoi, and J.~Zenihiro,
Phys. Rev. C \textbf{84}, 054308 (2011).

\bibitem{ACCC} V.I. Kukulin and V.M. Krasnopol'sky, 
                 J. Phys. A \textbf{10} L33 (1977);
               V.I. Kukulin, V.M. Krasnopol'sky, and M. Miselkhi,
                  Sov. J. Nucl. Phys. \textbf{29}, 421 (1979).

\bibitem{ACCC-book} V.I.~Kukulin, V.M.~Krasnopol'sky, and
J.~Hor\'{a}\v{c}ek, \textit{Theory of resonances: Principles and
Applications}, (Kluwer Academic Publishers, Dordrecht,
Netherlands, 1989), p.~219.


\bibitem{ACCC-Aoyama}  S.~Aoyama,
  Phys. Rev. C \textbf{68}, 034313 (2003).

\bibitem{Arai} K.~Arai,
  Phys. Rev. C \textbf{74} (2006) 064311.

\bibitem{Saito69} S. Saito,
  Prog. Theor. Phys. \textbf{40}, 893 (1968); 
                     \textbf{41}, 705 (1969);
  Prog. Theor. Phys.(Suppl.) \textbf{62}, 11 (1977).

\bibitem{Matsumoto03} T. Matsumoto, T. Kamizato, K. Ogata, Y.Iseri,
E. Hiyama, M. Kamimura, and M. Yahiro, Phys. Rev. C \textbf{68}, 064607
(2003).

\bibitem{Nakada} H.~Nakada, K.~Mizuyama, M.~Yamagami, and M.~Matsuo,
  Nucl. Phys. A \textbf{828}, 283 (2009).

\bibitem{Kamimura09} M. Kamimura, E. Hiyama, and Y. Kino,
  Prog. Theor. Phys. \textbf{121}, 1059 (2009).

\bibitem{Hiyama2012} E. Hiyama and M. Kamimura,
  Phys. Rev. A \textbf{85}, 022502 (2012);
  Phys. Rev. A \textbf{85}, 062505 (2012).

\bibitem{SW-force} E.W. Schmit and K. Wildermuth,
  Nucl. Phys. \textbf{26}, 463 (1961).

\bibitem{Kukulin84} V.I. Kukulin, V.M. Krasnopol'sky,
                    V.T. Voronchev, and P. B. Sazonov,
  Nucl. Phys. A \textbf{417}, 128 (1984).

\bibitem{Ajzenberg} F. Ajzenberg-Selobe, 
  Nucl. Phys. A \textbf{506}, 1 (1990).

\end{thebibliography}
\end{document}